\documentclass[10pt,journal]{IEEEtran} 
\usepackage[utf8]{inputenc}
\usepackage[T1]{fontenc}
\usepackage{algorithm}
\usepackage{amsmath}
\usepackage{amsthm}
\usepackage{amssymb}
\usepackage{graphicx}
\usepackage{setspace}
\usepackage{epstopdf}
\usepackage[font={normal}]{caption}
\usepackage{cite}
\usepackage{bm}
\usepackage{algpseudocode}
\usepackage{balance}
\usepackage{xcolor}
\usepackage{
  tabu                        
}
\usepackage{array,multirow,makecell}\setcellgapes{1pt}
\usepackage{comment}
\usepackage{setspace}
\usepackage{subfigure}
\usepackage{dsfont}
\graphicspath{{SecFigures/}}
\newcolumntype{R}[1]{>{\raggedleft\arraybackslash }b{#1}}
\newcolumntype{L}[1]{>{\raggedright\arraybackslash }b{#1}}
\newcolumntype{C}[1]{>{\centering\arraybackslash }b{#1}}
\makegapedcells
\makeatletter
\long\def\@IEEEtitleabstractindextextbox#1{\parbox{0.922\textwidth}{#1}}
\makeatother

\begin{document}

\title{A Blockchain-based Reliable Federated Meta-learning for Metaverse: A
 Dual Game Framework}
\author{Emna~Baccour,~\IEEEmembership{Member,~IEEE,}        
	Aiman~Erbad,~\IEEEmembership{Senior~Member,~IEEE,}	
    Amr~Mohamed,~\IEEEmembership{Senior~Member,~IEEE,}
	Mounir~Hamdi,~\IEEEmembership{Fellow,~IEEE,}
	and~Mohsen~Guizani,~\IEEEmembership{Fellow,~IEEE}
\thanks{E. Baccour, A. Erbad, and M. Hamdi are with the Division of Information and Computing Technology, College of Science and Engineering, Hamad Bin Khalifa University, Qatar Foundation, Doha, Qatar (email: ebaccourepbesaid@hbku.edu.qa; aerbad@ieee.org; mhamdi@hbku.edu.qa).}
\thanks{A. Mohamed is with the College of Engineering, Qatar University, Doha, Qatar (email: abensaid@qu.edu.qa; amrm@qu.edu.qa).}
\thanks{M. Guizani is with  the Machine Learning Department, Mohamed Bin Zayed University of Artificial Intelligence (MBZUAI), Abu Dhabi, UAE (email: mguizani@ieee.org).}
\thanks{This paper is published in IEEE Internet of Things Journal (2024)~\cite{10485381} and a part of it was published in IEEE Wireless Communications and Networking Conference (2024)~\cite{10571133}.}
}

\IEEEtitleabstractindextext{
\begin{abstract}
The metaverse, envisioned as the next digital frontier for avatar-based virtual interaction, involves high-performance models. In this dynamic environment, users' tasks frequently shift, requiring fast model personalization despite limited data. This evolution consumes extensive resources and requires vast data volumes. To address this, meta-learning emerges as an invaluable tool for metaverse users, with federated meta-learning (FML), offering even more tailored solutions owing to its adaptive capabilities. However, the metaverse is characterized by users heterogeneity with diverse data structures, varied tasks, and uneven sample sizes, potentially undermining global training outcomes due to statistical difference. Given this, an urgent need arises for smart coalition formation that accounts for these disparities. This paper introduces a dual game-theoretic framework for metaverse services involving meta-learners as workers to manage FML. A blockchain-based cooperative coalition formation game is crafted, grounded on a reputation metric, user similarity, and incentives. We also introduce a novel reputation system based on users' historical contributions and potential contributions to present tasks, leveraging correlations between past and new tasks. Finally, a Stackelberg game-based incentive mechanism is presented to attract reliable workers to participate in meta-learning, minimizing users' energy costs, increasing payoffs, boosting FML efficacy, and improving metaverse utility.  Results show that our dual game framework outperforms best-effort, random, and non-uniform clustering schemes - improving training performance by up to 10\%, cutting completion times by as much as 30\%, enhancing metaverse utility by more than 25\%, and offering up to 5\% boost in training efficiency over non-blockchain systems, effectively countering misbehaving users.
\end{abstract}

}
\maketitle
\IEEEdisplaynontitleabstractindextext
\IEEEpeerreviewmaketitle
\section{Introduction}
Recently, the concept of metaverse has gained considerable attention, driven by the COVID-19 pandemic's constraints on physical mobility and the interests of major tech companies \cite{survey3}. Through customizable 3D avatars, users can engage in activities mirroring real-life experiences within this shared virtual reality. The advancement of metaverse technology has substantially reduced the barriers between physical and digital worlds, aided by technologies like digital twins, blockchain, extended reality (XR), and AI algorithms \cite{survey4}. As this digital frontier evolves, high-performing AI models are required to ensure a seamless user experience and real-time rendering. However, a significant challenge arises from the scarcity of data within each user's individual context and the need for frequent model personalization due to the ever-shifting nature of tasks. These dynamics demand extensive resources, vast data volumes, and high learning time to train robust and adaptive models.

To overcome this limitation and enhance model performance, a collective training approach, Federated Learning (FL), is required, facilitating collaborative model refinement across decentralized devices. The integration of Federated Learning in the metaverse (FL4M) \cite{chen2023federated} promises enhanced data privacy and security, alongside decentralized machine learning that leverages user-generated data and computational resources for model building. Yet, in the metaverse, effective management of heterogeneous data is crucial to maintain data utility. In FL, data heterogeneity arises due to diverse client data characteristics, often exhibiting non-independent and non-identically distributed (non-IID) properties, affecting model accuracy. This heterogeneity is not limited to data characteristics; indeed, the metaverse has enormous number of users, objects, and interactions, which creates scalability challenges for traditional FL algorithms. Users may have different preferences, devices, and learning tasks, which can create challenges to adapt to the behavior of the avatars, pursue the ever-shifting nature of their tasks, and provide a more personalized experience, especially that FL mainly focuses on learning one ML task across multiple devices \cite{FL3}. Furthermore, users expect a responsive and adaptive virtual environment, which cannot be achieved through FL that requires long re-training. Addressing these multifaceted heterogeneity aspects is crucial for efficient FL solutions in the metaverse.

Federated Meta-Learning (FML), known as "learning to learn", emerges as a solution to tackle multi-task learning challenges \cite{FML}, while preserving the benefits of Federated Learning, such as data security and flexibility \cite{PervasiveAI}. Particularly, initialization-based meta-learning algorithms are well known for fast adaptation and strong generalization to new tasks \cite{ modelagnostic}. In other words, it enables a model to adapt swiftly to new tasks by using its prior experience from other tasks in real-time. As an illustrative example, in a vehicular metaverse, distinct users engage within similar virtual environments but with differing tasks and objectives, such as optimizing fuel efficiency and improving driving techniques. FML demonstrates its prowess by seamlessly accommodating these divergent tasks using different data and offering personalized virtual driving models catering to each user's unique needs. Additionally, FML considers evolving tasks within the metaverse, such as simulating snowy roads, efficiently personalizing previously generated models with minimal data samples, computation resources, and learning latency.



\begin{figure*}[h]
\centering
	\includegraphics[scale=0.6]{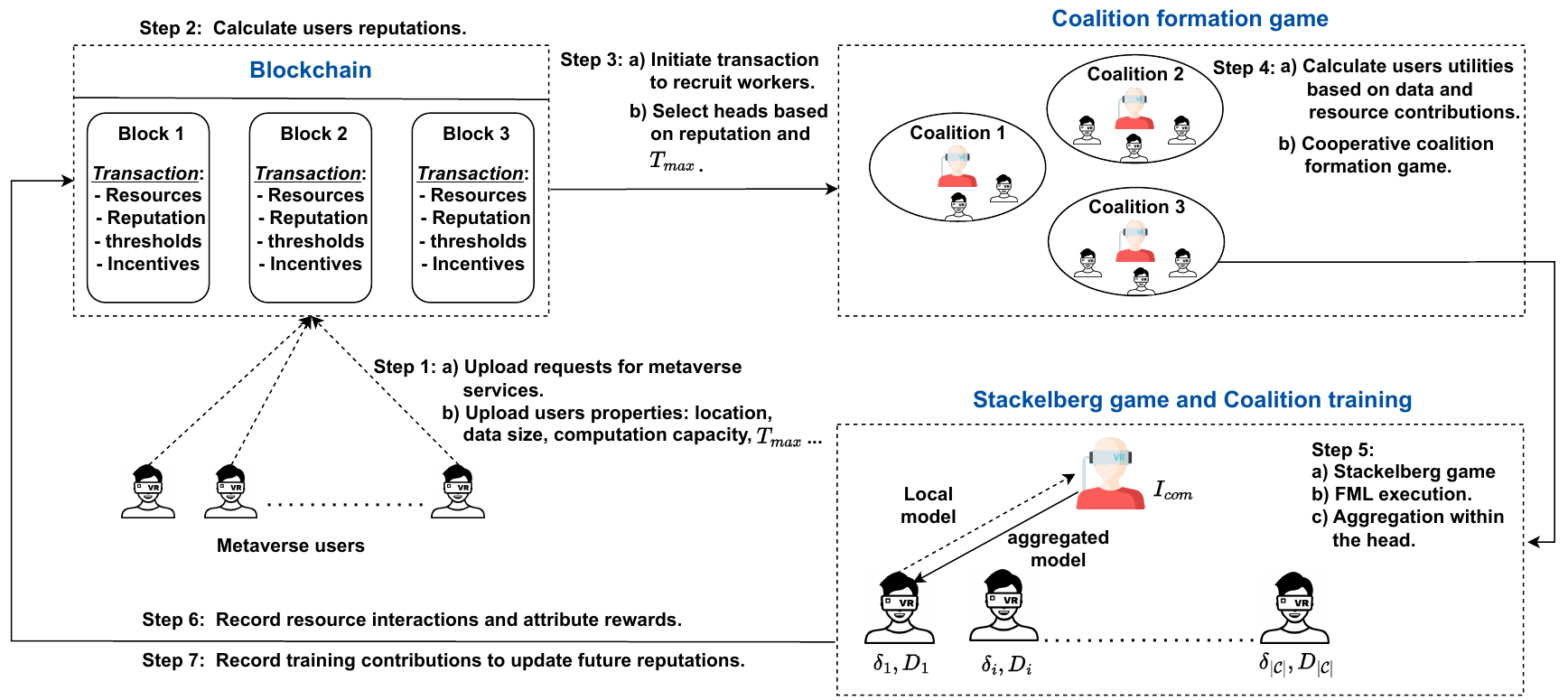}
	\caption{Metaverse dual game framework.}
	\label{sm}
\end{figure*}

However, despite its promising benefits, FML comes with new challenges in the metaverse. Specifically, three critical issues must be addressed. First, the metaverse has a vast number of heterogeneous users with diverse data structures, tasks, and sample sizes, making collective training inefficient due to unrelated tasks and leading to low convergence speed with random uniform device selection \cite{FML,meta-learning2}. Non-uniform device selection widely applied for FL \cite{PervasiveAI} face challenges in FML due to bias, statistical heterogeneity, and high-order information in stochastic gradients. Excluding meta-learners with minimal contributions, as proposed in \cite{meta-learning3}, is not feasible in the metaverse, where users actively seek participation to craft personalized experiences.  Therefore, a clustering mechanism should be wisely designed to group metaverse users based on their tasks and data similarity. Coalition formation games have been introduced in the context of FL to support Hierarchical Federated Learning (HFL) techniques \cite{coalitionFormation1} or to optimize communication resource efficiency \cite{commCost}. Nevertheless, the high diversity of tasks within the metaverse poses a challenge, making conventional clustering methods unsuitable for direct application. 
Secondly, some metaverse users may misbehave to selfishly increase their utility, resulting in a bad training experience. Hence, clustering mechanism should consider the presence of unreliable users to create a fair training circumstances among different coalitions in order to minimize the impact on each coalition's training performance. Third, users' devices may be unwilling to engage in metaverse services due to resource allocation concerns, energy consumption, and the absence of appropriate incentives. While resource allocation and incentive mechanisms for FL have been the focus of extensive research, particularly for metaverse \cite{HFedMS,10304077}, FML remains underexplored.

In this paper, to the best of our knowledge, we are the first to design a framework that manages the usage of federated meta-learning to empower metaverse services. Our novel distributed computing framework integrates FML techniques with blockchain technology \cite{blockchain,10047856} to address the unique challenges of the metaverse previously discussed and pave the way for efficient and personalized learning strategies. The framework introduces a dual game mechanism consisting of a cooperative coalition formation game for clustering metaverse meta-learners and a Stackelberg game for resource allocation and incentives, which are novel approaches not explored before in the context of FML. On one hand, the coalition formation game clusters metaverse meta-learners based on their tasks similarities, contributions and reputations, addressing the inefficiencies caused by statistical heterogeneity and ensuring average reliability of each coalition. On the other hand, the Stackelberg game serves as an incentive mechanism to ensure fair payoff distribution among metaverse users within each coalition according to their contribution. The framework operates in steps process, starting with users requesting personalized meta-models from a metaverse service provider and concluding with the update of different models and reputations of users after each FML round. Throughout training rounds where both games interact, our framework enables efficient meta-training, enhances users contributions and payoffs, and reduces energy costs while promoting collaboration and fairness among them. Blockchain technology underpins the framework, to securely manage reputations and incentives, track participation, and ensure transparent interactions among metaverse entities.
Our contributions are:
\begin{itemize}
\item 
    We introduce a novel blockchain-based reliable federated meta-learning framework designed to support immersive user experiences in the metaverse. Our framework leverages a unique dual game mechanism to address the challenges of tasks heterogeneity and prompt personalization.
    \item We propose a cooperative coalition formation game to cluster metaverse users based on their computation contribution, model embedding similarity, and overall reputation, which is partly dependent on past contributions to similar tasks. This proposed approach effectively addresses the inefficiencies caused by statistical diversity in the metaverse.
    \item We introduce a novel method that quantifies the impact of users on model performance and tasks completion time as contribution. On these bases, a reputation management scheme is designed based on historical contributions and potential contributions to present tasks, leveraging correlations between past and new tasks.
    \item 
    We formulate a Stackelberg game to incentivize reliable  users to participate in metaverse FML services and improve system utility. This game-theoretic approach ensures optimal resource allocation and fair payoff distribution among FML participants within each coalition.
    \item We present an algorithm, namely Game-Theoretic Federated Meta-Learning (GFML) for the metaverse, that leverages the interaction between the dual games.
    \item Simulation results show that our proposed  framework in a metaverse system guarantees better learning performance and resilience against malicious metaverse users. Additionally, users can obtain higher profits compared to the best-effort, random and non-uniform clustering schemes. 
\end{itemize}
This paper is organized as follows: Section \ref{section:RelatedWorks} provides an overview of existing literature on federated learning and federated meta-learning. In Section \ref{section:Framework}, we detail the framework and system model for our proposed blockchain-based reliable federated meta-learning in the metaverse. The development of the reputation model and the coalition formation game are introduced in Section \ref{section:CoalitionFormation}. The incentive mechanism based on the Stackelberg game is formulated in Section \ref{section:Stackelberg}. An evaluation of the performance of our reliable metaverse framework is presented in Section \ref{section:evaluation}. Finally, Section \ref{section:conclusion} concludes the paper.

\section{Related works}\label{section:RelatedWorks}
The metaverse has gained significant attention, driving research into integrating advanced technologies and AI techniques, such as digital twin~\cite{survey4}, XR~\cite{survey3}, blockchain~\cite{9881813}, and semantic learning~\cite{10208153}. Federated learning has particularly garnered interest for its ability to ensure data privacy and leverage user resources in the metaverse~\cite{chen2023federated}. For example, Zhou et al.\cite{10304077} proposed an optimization framework for FL in mobile augmented reality within the metaverse, focusing on minimizing energy consumption and completion time while improving model accuracy. Kang et al.~\cite{9881813} explored a privacy-preserving FL framework for the industrial metaverse using blockchain. However, these works primarily focused on resource and time performance, overlooking the critical aspect of data heterogeneity among learners. Zeng et al.\cite{HFedMS} presented HFedMS, an FL system for the industrial metaverse that addresses data heterogeneity and learning forgetting through clients regrouping and semantic compression and compensation. Most of federated learning efforts in the metaverse, including HFedMS, aim to address the core issue which is data heterogeneity. Nevertheless, the metaverse's heterogeneity extends beyond data characteristics and user preferences, encompassing system and protocol variations that challenge traditional FL. Furthermore, the dynamic and personalized nature of the metaverse, with users having evolving and diverse tasks, poses additional challenges for FL algorithms, which primarily focus on learning a single task across devices.
Efficiently addressing these multifaceted heterogeneity issues is crucial for enhancing collaborative training performance within the metaverse's unique ecosystem, which is the focus of this paper. Federated Meta-Learning offers a promising solution by enabling fast adaptation to new tasks and heterogeneous users, increasing learning accuracy for diverse multi-task environments, and allowing diverse metaverse users to handle a wide range of scenarios and contexts. Despite its advantages, FML has not been integrated or adapted for the metaverse, leaving user-specific needs unmet. To our knowledge, we are the first to propose the integration and adaptation of FML techniques to cater to the tasks-heterogeneity of users in the metaverse dynamic environment.

Recent literature shows growing interest in FML, with efforts focusing only on enhancing model convergence and accuracy through optimized loss functions and averaging methods for general applications.  Among these mainstream works, Chen et al. introduced FedMeta, an FML method enhancing FL and FedAvg through a Model-Agnostic Meta-Learning (MAML) algorithm \cite{FML1} for high statistical and systematic challenges. Jiang et al. explored the relationship between MAML and FedAvg, demonstrating improved and stable personalized performance offered by FML \cite{FML2}. Lin et al. analyzed FML convergence considering adversarial data and strongly convex loss functions \cite{FML3}, while Fallah et al. confirmed FML's convergence in non-convex scenarios with stochastic gradients \cite{FML}. The FML algorithms described above frequently face challenges with slow convergence and inefficient communication due to random device selection. Non-uniform device selection has been proposed in meta-learning \cite{meta-learning3}, which involves the exclusion of users with minimal contributions to the training. This approach is impractical in metaverse, as users actively seek participation to create personalized experiences.  Clustering all active users looking for a customized experience according to their tasks and context similarities, is not explored in the literature. Our paper will not focus on enhancing loss and averaging performance as is common in existing works, where well-established techniques prevail. Instead, we aim to bridge a significant gap by proposing a novel method for dynamic clustering tailored to the unique needs of the metaverse.

Another critical gap in FML literature is the lack of comprehensive strategies for effective resource allocation in distributed device environments \cite{meta-learning3}. Unlike FML, Federated Learning has seen considerable research dedicated to enhancing clustering, users selection, and resource allocation \cite{commCost,coalitionFormation1,8832210,ZHU2023817,9917252}. He et al. propose a three-stage Stackelberg game-enabled clustered federated learning approach for heterogeneous UAV swarms, where game-theoretic framework is employed to optimize learning, resource consumption, and incentive allocation~\cite{commCost}. Ng et al. also propose a reputation-aware hedonic coalition formation approach for efficient serverless hierarchical FL, enabling participants to form stable coalitions based on their preferences and reputations \cite{coalitionFormation1}. Kang et al. introduce an incentive mechanism that combines reputation and contract theory to ensure reliability and honest user participation in FL \cite{8832210}. Zhu et al. present a dynamic incentive and reputation mechanism to promote energy efficiency and motivate user participation in FL for 6G networks \cite{ZHU2023817}.  Allahham et al. develop a Stackelberg game approach to incentivize learners in multi-orchestrator mobile edge learning environments \cite{9917252}. It is worth noting that several studies have explored resource allocation techniques based on hierarchical and Stackelberg game frameworks and AI techniques in edge and cloud computing systems, such as the works in \cite{9722568,9206115}. These approaches consider the dynamic nature of users and their influence on resource allocation.

While FL has been extensively researched, FML remains underexplored, and the clustering and selection techniques developed for FL cannot be directly applicable to FML due to higher-order information and biased stochastic gradients. Our proposed framework introduces a novel clustering technique tailored for FML and metaverse relying on embedding similarity, historical contributions, and potential contributions to present tasks, leveraging correlations between past and new tasks. Additionally, we have designed a novel dual-game framework that utilizes coalitions to enhance training efficiency and a Stackelberg game to incentivize users to contribute their resources and improve the utility of the metaverse.

\begin{table*}[t]
	\centering
 \small
	\caption{List of key Notations.}
	\label{tab1}
	\begin{tabular}{C{2cm}|L{5.8cm}||C{2cm}|L{5.8cm}}
		\hline   \textbf{Symbol} & \textbf{Description} & \textbf{Symbol} & \textbf{Description}\\
		\hline  
         $\mathcal{N}$& Set of metaverse meta-learners   &$\zeta$  & Effective capacitance \\
	   $N^a$&Number of active metaverse users& $c^{com}_{j,i}$& Computation cost of user $i \in \mathcal{C}_j$\\
         $N^p$&Number of passive metaverse users& $T^{comp}_{j,i}$  & Computing latency \\
         $\mathcal{D}_i$&Dataset of user $i$&$c^{comp}_{j,i}$  &Communication cost \\
          $T^i_{max}$& Tolerated task completion time of $i$  &$T^{comm}_{j,i}$  &Transmission time \\
          $\underline{\delta^i}$, $\overline{\delta^i}$& Max and min CPU frequency of $i$  & $B_{j,i}$ & Mean transmission rate of the user $i$\\
       $\tau$& Number of local iteration   & $U^{MML}_{j,i}$ & Profit of the user $i\in\mathcal{C}_j$ \\
        $\theta$& Weights of the DNN & $U^{MSP}_j$ &Utility of the MSP from the coalition $\mathcal{C}_j$ \\
        $\beta$ & Step size  &$S_{j,i}$  & Similarity of data distribution  \\
        $\alpha$& Meta-learning rate   &$H(.)$  &Reputation function \\
         $M$  & Number of coalitions  & $R_{j,i}$ &Reputation of user $i\in\mathcal{C}_j$ \\
          $\mathcal{H}$& Set of heads  & $\gamma$  & Default reputation utility \\
           $\Pi$&Coalitional partition   & $R^{global}_{j,i}$  & Global reputation   \\
           $\mathcal{C}_j$& Coalition $j$   & $R^{task}_{j,i}$ &Task-specific reputation \\
          $I_{rep}$, $I_{comp}$& Reputation and competition incentives   &$\phi$  &Reputation weighting parameter \\
          $R_{th}$& Reputation threshold   &$\mu$  &Utility parameter \\
           $\rho$& Unit computation cost& $\lambda$ & Time declining parameter \\
            $c_{j,i}$& CPU to update the model with 1 sample &$w^k_i$  &Embeddings of $i$ at round $k$ \\
            $\vartheta^k_{j,i}$& Contribution of $i$ to the meta-training&$u^k_{i,j}$&Contribution of $i$ to the reduction of loss\\
		\hline 
	\end{tabular}
\end{table*}
\section{Framework and System model}\label{section:Framework}
Fig. \ref{sm} presents the federated meta-learning process in the metaverse, including the coalition formation phase and the meta-learning execution phase guided by the stackelberg game. 
\subsection{Framework}
Our framework introduces a novel integration of federated meta-learning within the metaverse, leveraging blockchain technology for decentralized management of interactions among metaverse Meta-Learners (MMLs). At its core, the framework is designed to ensure reliable and real-time personalized experiences through a dual game managing metaverse users and their contributions to the FML process.  Specifically, the process unfolds in seven sequential steps, beginning with MMLs requesting meta-models from a metaverse Service Provider (MSP) and concluding with the MSP updating the requested models and reputations of MMLs post each FML round. Central to this process is the cooperative coalition formation game clustering MMLs based on utility maximization, and a Stackelberg game framework for optimal task allocation and incentives distribution. Blockchain technology underpins the framework, managing the reputations of MMLs through a decentralized and tamper-proof mechanism, securing transactions between MMLs and the MSP, and ensuring that incentives are distributed fairly and transparently. It also records all interactions, formed coalitions and MMLs' contributions, thereby providing an immutable and trusted history that can be referenced for future coalition formation, tasks allocation, incentives distribution, and reputation updates. Details about this process are introduced as follows:

\textit{Step 1:} A set $\mathcal{N}$ of metaverse meta-learners request meta-models from the MSP through the blockchain layer. Each MML looks for a good model initialization, i.e., meta-model, that can be quickly re-fined to a high-performing model via one or few gradient descent steps $\tau \geq 1$, in order to be used for the personalized tasks and potential new learning tasks. Furthermore, each MML $i \in \mathcal{N}$ has a labeled private dataset $\mathcal{D}_i=\{x^i_j,y^i_j\}_{j=1}^{|D_i|}$, where $(x^i_j,y^i_j) \in \mathcal{X} \times \mathcal{Y}$ is a data input following an underlying distribution $P_i$ with $x^i_j$ denoting the sample and $y^i_j$ denoting the label. MMLs might opt to join the FML process, offering their data and computing resources in exchange for incentives (Active MMLs), or they might choose to acquire a pre-trained model and perform only the personalization phase (Passive MMLs). Hence, the set $\mathcal{N}$ is defined as $\mathcal{N}=\{1,...,N^a\} \cup \{1,...,N^p\}$, where $N^a$ is the number of active MMLs and $N^p$ is the number of passive MMLs. It is worth noting that each active learner $i$ submits its request with its tolerated task completion time $T^i_{max}$, maximum and minimum CPU frequency, i.e., $\underline{\delta^i}$ and $\overline{\delta^i}$, that it can offer, its location, and its data characteristics. 

\textit{Step 2:} 
To evaluate the reliability of active MMLs, the blockchain system calculates composite reputation scores derived from the historical contributions of various users in previous tasks. Miners play an essential role in this process by being incentivized to safely record the users' interactions with the metaverse and calculate their composite reputations when needed. 

\textit{Step 3:} 
The MSP initiates a blockchain transaction to recruit the MMLs willing to serve for the learning tasks in the metaverse and choose among them a set $\mathcal{H}=\{h_1,...,h_{M}\}$ of $M$ heads with lower task completion constraints and superior blockchain-recorded reputations. The heads serve as aggregators for the FML tasks. The transaction contains the requirements of the MSP, such as the reputation threshold $\mathcal{R}_{th}$, and the incentives $I_{rep}$ and $I_{comp}$ for reputation and competition contributions, respectively. It also includes the initial model parameters, such as the weights $\theta$ of the Deep Neural Network (DNN) in addition to the number of local computational epochs $\tau$ that should be processed by any active MML.

\textit{Step 4:} Due to data heterogeneity and task objectives, MMLs are clustered into $M$ coalitions, i.e., $\Pi=\{\mathcal{C}_1,...,\mathcal{C}_M\}$, led by the selected heads using a cooperative coalition formation game. In cooperative games, the MMLs which are self-interested and focus on maximizing their own utilities, exhibit preferences regarding the clusters they wish to join
based on their statistical similarity to the head, the incentives they can gain, and the computation energy cost. As such, their utilities are influenced solely by the members of the coalitions they are part of. Throughout the game, coalitions may need to be dynamically adjusted. The miners are responsible in updating the reputation scores of each participant, recalculating utilities, and ensuring that new coalitions remain beneficial to all members involved. Once a coalition is formed, the miners record on the blockchain the final coalitions in addition to the combinations that did not lead to a Nash stability.

\textit{Step 5:} The MSP allocates the FML training tasks to the active MMLs within each coalition, using a Stackelberg game framework. In this game setup, the MSP acts as the leader across all coalitions, thereby dictating the reward strategy. Simultaneously, the MMLs, comprising both the members and the heads of the coalitions, function as followers who fine-tune their computing speed strategies. This arrangement highlights that distinct games occur within each coalition, despite the MSP's overarching the leadership across all games. The federated meta-learning aims to solve the following problem \cite{FML}:
\begin{equation}
	\label{eq:1}
   \underset{\theta \in \mathbb{R}^d} \min F(\theta)=\frac{1}{N^a} \underset{i \in N^a}{\sum} f_i(\theta-\beta \nabla f_i(\theta)),
\end{equation}
where $\beta$ is the step size, and $f_i(\theta)=\mathbb{E}_{(x,y)\backsim P_i}[l_i(\theta;x,y)]$ denotes the expected loss function over the data distribution of the MML $i$. We note that $l_i(\theta;x,y)$ is the loss function of the model parameters $\theta$ indicating the prediction errors. Similarly to the FL, the FML addresses the problem in (\ref{eq:1}) through two iterative steps, namely local updates and a global aggregation:
\begin{itemize}
    \item Local updates: At each global round $k$, each head $h_j$ sends the current model $\theta^k_j$ to its coalition members. Next, each MML $i$ personalizes the received model to its own task and updates it based on its meta-function $F(\theta)= f_i(\theta-\beta \nabla f_i(\theta))$ by running locally $\tau$ steps of the stochastic gradient descent. In particular, these local updates generate a local sequence $\{\theta_{j,i}^{k,t}\}_{t=0}^{\tau}$,where $\theta_{j,i}^{k,t}$ presents the local model of the MML $i$ participating in the coalition $j$ and running the $t$-th step of the local update in the round $k$, and $\theta_{j,i}^{k,0}=\theta_{j}^{k}$ \cite{FML}:
    \begin{equation}
	\label{eq:2}
   \theta_{j,i}^{k,t+1}=\theta_{j,i}^{k,t} - \alpha \Tilde{\nabla} F_i(\theta_{j,i}^{k,t}), \forall~0 \leq t \leq \tau-1.
\end{equation}
$\alpha>0$ denotes the meta-learning rate and  $\Tilde{\nabla} F_i(\theta_{j,i}^{k,t})$ is the stochastic gradient for all local iterates and it is calculated using independent batches $\mathcal{D}^t_{j,i}$, $\mathcal{D}^{t'}_{j,i}$,and  $\mathcal{D}^{t''}_{j,i}$~\cite{FML}\footnote{ For sake of clarity, we simplify  $\theta_{j,i}^{k,t}$ to $\theta$ and $\mathcal{D}^t_{j,i}$, $\mathcal{D}^{t'}_{j,i}$, $\mathcal{D}^{t''}_{j,i}$ to $\mathcal{D}_{j,i}$, $\mathcal{D}'_{j,i}$, $\mathcal{D}''_{j,i}$.}:
\begin{equation}
\footnotesize
	\label{eq:3}
 \begin{split}
  \Tilde{\nabla} F_{j,i}(\theta)=(I-\beta\Tilde{\nabla}^2 f_{j,i}(\theta,\mathcal{D}''_{j,i}))\Tilde{\nabla}f_{j,i}(\theta-\beta\Tilde{\nabla}f_{j,i}(\theta,\mathcal{D}_{j,i}),\mathcal{D}'_{j,i})
  \end{split}
\end{equation}
We note that, for any batch $\mathcal{D}_{j,i}$, $\Tilde{\nabla}f_{j,i}(\theta,\mathcal{D}_{j,i})$ and $\Tilde{\nabla}^2f_{j,i}(\theta,\mathcal{D}_{j,i})$ are the unbiased estimates of $\nabla f_{j,i}(\theta)$ and $\nabla^2f_{j,i}(\theta)$, respectively:
 \begin{equation}
 \begin{split}
	\label{eq:4}
 \Tilde{\nabla}f_{j,i}(\theta,\mathcal{D}_{j,i})=\frac{1}{|\mathcal{D}_{j,i}|} \underset{(x,y)\in\mathcal{D}_{j,i}}{\sum}\Tilde{\nabla}l_{j,i}(\theta;x,y)\\  
 \Tilde{\nabla}^2f_{j,i}(\theta,\mathcal{D}_{j,i})=\frac{1}{|\mathcal{D}_{j,i}|}\underset{(x,y)\in\mathcal{D}_{j,i}}{\sum}\Tilde{\nabla}^2l_{j,i}(\theta;x,y)
 \end{split}
\end{equation}
However, $ \Tilde{\nabla} F_{j,i}(\theta)$ is a biased estimate of $ \nabla F_{j,i}(\theta)$ as it is expressed as a  stochastic gradient that contains another stochastic gradient inside. This fact makes FML different from FL where the local updates are done using unbiased gradient estimates.
\item Global aggregation: After local updates, each active MML returns its local model parameters, i.e., $\theta^k_{j,i}=\theta^{k,\tau-1}_{j,i}$, to the corresponding head $h_j$, which then updates the global model received from its coalition members, i.e.,

     \begin{equation}
 \begin{split}
	\label{eq:5}
 \theta_j^{k+1}=\frac{1}{|\mathcal{C}_j|}\underset{i\in \mathcal{C}_j}{\sum}\theta^k_{j,i}
 \end{split}
\end{equation}
After the end of the training and finding an initial shared model that performs well, each MML updates it with respect to its task and local dataset, possibly by one or a few steps of gradient descent using its own loss function.
\end{itemize}
To summarize, The strength of FML lies in its ability to preserve FL's benefits, while capturing the difference between users. Both active and passive MMLs can use the solution of this problem as a starting point and make slight updates based on their data and tasks. Essentially, MMLs use the initial meta-model that is trained to yield a promising individualized model for each user after just one or a few local gradient steps.

\textit{Step 6:} Blockchain records resource interactions between the MSP and MMLs at each round, enabling users to receive fitting incentives from the MSP, which could include tokens, credits, or premium services access.

\textit{Step 7:} Upon completing each FML round, the MSP updates the reputation of each MML on the blockchain, with respect to its respective coalition.
\subsection{System model}
\begin{figure}[h]
\centering
\hspace{-8mm}
	\includegraphics[scale=0.57]{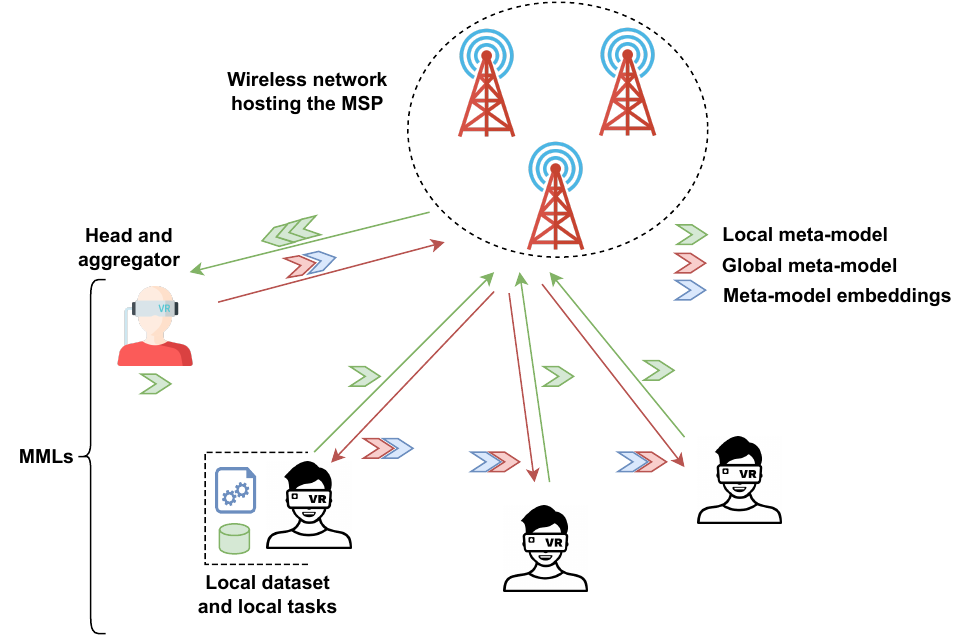}
	\caption{System model of the FML for metaverse over a wireless network with multiple MML devices.}
	\label{architecture}
\end{figure}

We consider a wireless multi-user system, where a set  of $N^a$ MML devices collaborate in coalitions to carry out federated meta-learning aided by the wireless network, as illustrated in Fig. \ref{architecture}.
Each global round consists of two phases: the computation phase and the communication phase. In the computation phase, each MML device $i \in \mathcal{C}_j$ downloads the meta-model embeddings of heads to calculate its utility and choose its coalition, as will be detailed in the subsequent section. Once the coalitions are formed, the MML receives the current global model from the head and computes its local model using its own dataset. During the communication phase, the MML devices send their local models to the respective head via wireless nodes. After that, each head aggregates the received models and starts the following round. For the sake of simplicity, we omit the consideration of downlink communication due to the asymmetric uplink-downlink nature of wireless networks. This implies that the transmission power and downlink communication bandwidth, typically stronger at the server (e.g., a base station), make the downlink time negligible compared to uplink data transmission time \cite{meta-learning3}. 

Considering the learning process within a coalition consumes computational energy, the computing cost of each MML $i$ can be depicted as follows\footnote{As our primary focus is resource allocation and incentives in each round $k$, we simplify our notation by omitting the subscript $k$ throughout this section.}: 
 \begin{equation}
 \begin{split}
	\label{eq:6}
c^{comp}_{j,i}=\rho \tau c_{j,i} |\mathcal{D}_{j,i}| \zeta (\delta_{j,i})^2,
 \end{split}
\end{equation}
where $\rho$ is the unit computation cost and $\tau$ denotes the number of iterations during one round $k$. We denote $c_{j,i}$ as the CPU cycles to update the model with one sample, and $|\mathcal{D}_{j,i}|$ as the batch size used in local update. Therefore, $c_{j,i} |\mathcal{D}_{j,i}|$ is the number of CPU cycles required to run one iteration. $\zeta$ represents the effective capacitance parameter of the computing chipset of MML $i$. Finally, $\delta_{j,i}$ is the computation capability of the device $i$, which is determined by the clock frequency of the CPU. Furthermore, we can express the time cost of the computing phase for a device $i$ in the iteration $k$, as follows:
 \begin{equation}
 \begin{split}
	\label{eq:7}
T^{comp}_{j,i}=\frac{\tau c_{j,i} |\mathcal{D}_{j,i}|}{\delta_{j,i}}.
 \end{split}
\end{equation}
Moreover, the energy consumption to transmit the updated model is expressed as follows \cite{commCost}:
 \begin{equation}
 \begin{split}
	\label{eq:8}
c^{comm}_{j,i}= a (\frac{\delta_{j,i}}{1-\epsilon})^2+b (\frac{\delta_{j,i}}{1-\epsilon})+z,
 \end{split}
\end{equation}
where $a$, $b$, and $z$ are the coefficients of communication costs, $\epsilon$ indicates the loss rate during transmission, and $(\frac{\delta_{j,i}}{1-\epsilon})$ presents the energy consumed accounting for transmission losses. The energy consumed for the communication between the MML and the coalition head is negligible compared to the energy consumed for computational tasks. Hence, for simplicity, we assume that $ c^{comm}_{j,i} \ll c^{comp}_{j,i}$. The transmission time is presented by:
 \begin{equation}
 \begin{split}
	\label{eq:9}
T^{comm}_{j,i}=\frac{W}{B_{j,i}},
 \end{split}
\end{equation}
where $W$ is the size of the local model pramaters, and $B_{j,i}$ is the mean transmission rate from user $i$ to the wireless node. 
\section{Reliable heads selection and coalition formation}\label{section:CoalitionFormation}
In order to incentivize users to join the meta-learning coalitions actively, the metaverse service provider attributes rewards to workers that contribute to metaverse services in exchange for their consumed resources.
\subsection{Profit function of metaverse meta-learners}
Rewards can be categorized into basic or reputation incentives $I_{rep}$ and competition incentives $I_{comp}$. The MMLs that participate in the same meta-learning coalition can receive an equal reward from $I_{rep}$ based on the average reputation of the group. Furthermore, the MMLs compete to maximize their overall reward by offering their local data. Specifically, MML $i$ with a larger dataset and a data distribution $P_i$ similar to the head can earn higher competition profits from $I_{comp}$. Additionally, each MML adopts the dynamic voltage scaling technology, enabling adaptive modification and management of their processing speed to obtain a higher profit. On the other hand, participating with the computing resources costs the MML energy. Hence, each participant's utility is calculated as the difference between the acquired rewards and the consumed energy. A rational MML will keep its utility function positive. It is also imperative that each user ensures their task completion time does not exceed the tolerated limit ($T^j_{max}$) required by the head $h_j$. On these bases, the profit or utility function of the MML $i \in \mathcal{C}_j$ is formulated in eq. (\ref{eq:10}).

\begin{equation}
\label{eq:10}
\begin{split}\raisetag{\baselineskip}
U^{MML}_{j,i}=(S_{j,i}+(\frac{\delta_{j,i}-\underline{\delta_j}}{\overline{\delta_j}-\underline{\delta_j}})\times(\frac{|\mathcal{D}_{j,i}|}{\underset{k\in \mathcal{C}_j}{\sum |\mathcal{D}_{j,k}|}}))\times I^j_{comp}\\ + \frac{\underset{k\in \mathcal{C}_j}{\sum H(\mathcal{R}_{j,i}))}}{|\mathcal{C}_j|} \times I_{rep} - c^{comp}_{j,i} - c^{comm}_{j,i}\qquad\qquad\qquad\qquad\\\\
\text{S.t. }~~~ \underline{\delta_j}=min(\underline{\delta^{i}},~\forall i \in \mathcal{C}_j)\qquad\qquad\qquad\qquad\\
\overline{\delta_j}=max(\overline{\delta^{i}},~\forall i \in \mathcal{C}_j)\qquad\qquad\qquad\qquad\\
 \end{split}
\end{equation}
where $I^j_{comp}$ is the competition incentives offered by the MSP to the coalition $\mathcal{C}_j$, $|\mathcal{C}_j|$ is the number of MMLs participating in the meta-learning task of the coalition $\mathcal{C}_j$, $\delta_{j,i}$ is the computing frequency of the MML $i$, and $|\mathcal{D}_{j,k}|$ is its data size.
$H(.)$ is the reputation function of the active MMLs, which is defined as \cite{Reliable_FL}:
\begin{equation}
	\label{eq:11}
H(\mathcal{R}_{j,i})= \Bigg\{
    \begin{array}{ll}
       \gamma+(1-\gamma)log(1-\mu)& \mbox{if } \mathcal{R}_{th}\leq \mathcal{R}_{j,i}\\
        \gamma e^{(\mathcal{R}_{j,i}-\mathcal{R}_{th})} & \mbox{if }  \mathcal{R}_{j,i} < \mathcal{R}_{th}
    \end{array}
\end{equation}
$\gamma$ is the default reputation utility, $\mathcal{R}_{th}$ is the reputation threshold specified by the metaverse service provider, and $\mathcal{R}_{j,i}$ is the reputation of the MML $i$ after joining $\mathcal{C}_j$. Furthermore, $\mu$ is expressed as follows \cite{blockchain1}:
\begin{equation}
\begin{split}
\mu=\frac{(e-1)(\mathcal{R}_{j,i} - \mathcal{R}_{th})}{(\overline{\mathcal{R}^{j}}-\mathcal{R}_{th})}\qquad\qquad\\
\text{S.t. }~~~ \overline{\mathcal{R}^{j}}=max(\mathcal{R}_{j,k},~\forall k \in \mathcal{C}_j)\qquad\qquad
\end{split}
\end{equation}
The reputation function $H(.)$ implies that when the reputation of an MML $i$ is lower than the reputation threshold, $H(.)$ decreases sharply; otherwise, $H(.)$ noticeably increases. Finally, $S_{j,i}$ is an index that assesses the similarity of data distribution between the MML $i$ and the head of the coalition $h_j$. Specifically, the metaverse environment is composed of users with different data structures and various tasks. To avoid random grouping of users and unstable global training performance caused by statistical heterogeneity, we encourage the collaboration between MMLs with similar data and tasks. Hence, we propose a statistical strategy to increase the utility of users with higher data similarity to its head. The $S_{j,i}$ index compares the embeddings \cite{embeddings} $\omega^k_j$ and $\omega^k_i$ of both head $h_j$ and MML $i$ derived from the initial model of the head in the current round $k$. The embeddings  are compared through the cosine similarity, as follows:
\begin{equation}
\begin{split}
S_{j,i}\triangleq\frac{<\triangle\omega^k_i,\triangle\omega^k_j>}{\parallel\triangle\omega^k_i\parallel\parallel\triangle\omega^k_j\parallel}.
\end{split}
\end{equation}
An embedding is a relatively low-dimensional space into which high-dimensional vectors (e.g., images) are transformed. In machine learning, embeddings are the output of \textit{Embedding layer}, which is the last layer before the head of the model. Embeddings function as task-specific dictionaries that capture some of the semantics of the inputs by placing semantically similar inputs close together in the embedding space. Thus, the similarity between two embeddings can provide insight into the similarity between the two tasks to be learned and, therefore, can help to proactively select the relevant metaverse users that can be trained together led by their head. The embedding can be obtained by a simple forward pass through the model with a reduced complexity. Moreover, it is considered a low-dimensional output. Hence, we will neglect its generation and transmission costs.

The similarity index $S_{j,i}$ can be calculated by comparing the embeddings of two samples or two small batches of data for higher accuracy. Since the model is gradually fine-tuned and metaverse users potentially generate more data for task training over time, $S_{j,i}$ is re-calculated at the beginning of each $\chi$ global round. When $\chi$ is equal to 1, the embeddings are calculated at each round to capture the updates periodically. The head broadcasts its embedding to all the coalition members, enabling the evaluation of their individual utilities. It is worth mentioning that the embeddings are low-dimensional data and less invasive compared to the input data. We also highlight the importance of selecting high-reputation heads in order to guarantee the integrity of the broadcasted embeddings. We finally note that the MMLs do not communicate with each other as their computation decisions are solely driven by their own utilities, regardless of the consequences on other users.
\subsection{Reputation model based on contribution}
In metaverse environment, each MML requests to acquire multiple models and participate in various meta-learning tasks. Hence, it is necessary for the MSP to comprehensively consider the long-term contributions of current users and their contributions in historical similar meta-learning tasks, especially as their data undergo continual changes. Additionally, it is impractical to assume that workers will always remain honest; some may maliciously attempt to manipulate their contributions to gain some advantages, such as joining specific coalitions by falsifying the reported similarity to the head's data,  misreporting their computational efforts to reduce energy consumption while still claiming high incentives, or disturbing the training to deteriorate the system performance.

\subsubsection{Contribution quantification of metaverse meta-learners}
The contribution of a given MML $i$ during the global round 
$k$ is determined by two factors: its contribution to model training and its task completion time. Accordingly, this contribution is quantified as:

 \begin{equation}
 \label{eq:contribution}
 \begin{split}
\vartheta^k_{j,i}=(u^k_{j,i}+1) \times log(1+T^j_{max}-T^{comp}_{j,i}-T^{comm}_{j,i}),
 \end{split}
\end{equation}
where $u^k_{j,i}$ is the contribution of the user $i$ to the reduction of the round global loss in the coalition $\mathcal{C}_j$ using  its local dataset $\mathcal{D}_{j,i}$ and gradient norm. We define the contribution $u^k_{j,i} \in [-1,1]$ as \cite{meta-learning3}:
 \begin{equation}
 \begin{split}
u^k_{j,i}=\sum_{t=0}^{\tau-1} \parallel \Tilde{\nabla} F_{j,i}(\theta_{j,i}^{k,t})\parallel^2-2(1+\frac{1}{\sqrt{|\mathcal{D}_{j,i}|}})\parallel \Tilde{\nabla} F_{j,i}(\theta_{j,i}^{k,t})\parallel
 \end{split}
\end{equation}
We note that the contribution $\vartheta^k_{j,i}$ is equal to 0 when MML $i$ has the lowest contribution to the training, as indicated by $u^t_{j,i}=-1$, and when it has the longest task completion time. Using this quantified contribution, we present a reputation mechanism to assess long-term user behavior. This reputation system takes into account both the user's global and task-specific reputation scores.
\subsubsection{Global reputation score}
The global reputation captures the historical performance of active MMLs within the metaverse system, reflecting their cumulative contributions across various training rounds and coalitions. To give prominence to recent contributions, we introduce contribution freshness metric. This is essential as a user's data and integrity can change over time. Furthermore, an honest user may be hijacked maliciously or may falsify its contribution (e.g., quality of data and computation resources). Thus, the global reputation is expressed as follows:
 \begin{equation}
 \label{eq:globalRep}
 \begin{split}
\mathcal{R}^{global}_{j,i}=\sum_{k=1}^{K} \lambda^k \vartheta^k_{j,i}
 \end{split}
\end{equation}
where $\lambda \in [0,1]$ is a constant. The term $\lambda^k$
represents a time-declining factor, ensuring that more recent contributions are given greater weight.
\subsubsection{Task-specific reputation score}
The task-specific reputation value refers to the performance of a user when participating in tasks related to the head $h_j$, which is formulated as follows:
 \begin{equation}
 \label{eq:taskRep}
 \begin{split}
\mathcal{R}^{task}_{j,i}=\sum_{k=1}^{K} \lambda^k(\mathds{1}_{r=j}~\vartheta^k_{r,i}+\mathds{1}_{r\neq j}~S_{j,r}~\vartheta^k_{r,i})
 \end{split}
\end{equation}
This reputation score comprises two parts:
\begin{itemize}
    \item \textit{Direct task-specific reputation:} For a task led by head $h_j$, this score captures the cumulative historical contributions of user $i$ in collaboration with the same head. It essentially represents a direct trust between the MML $i$ and its current head $h_j$.
    \item \textit{indirect task-specific reputation:} This denotes the aggregate historical contributions of user $i$ when working alongside other heads $h_r$ with $j\neq r$. Each contribution is adjusted by its similarity $S_{j,r}$ to the present task of $h_{j}$, thereby minimizing the influence of unrelated tasks. This score can be viewed as a recommendation derived from other aggregators.
\end{itemize}
Thus, the overall reputation value of an MML $i$ when joining a coalition $\mathcal{C}_j$ is as follows:
 \begin{equation}
 \label{eq:overall}
 \begin{split}
\mathcal{R}_{j,i}=\phi~\mathcal{R}^{global}_{j,i}+(1-\phi)\mathcal{R}^{task}_{j,i},
 \end{split}
\end{equation}
where $\phi \in [0,1]$ is a weighting parameter. If $\phi=1$, the reputation is general and not tied to any particular task. When $\phi=0$, the reputation solely reflects contributions similar to the current task. However, a user who has not collaborated with the current head or engaged in similar tasks will have a diminished reputation. This underscores the importance of considering both global and task-specific scores when evaluating overall contribution. In this way, active, reliable, and task-specialist MMLs can gain higher reputation values.

\subsection{Cooperative coalition game formulation}
The proposed federated meta-learning for metaverse involves creating disjoint coalitions, imposing the introduction of a cooperative coalition formation game. The two key requirements for classifying a coalitional game as a cooperative, also named hedonic game, are as follows \cite{coalitionFormation1}:

\textbf{Condition 1 (hedonic conditions).} A coalition formation game is classified as cooperative when it meets the following two conditions:
\begin{enumerate}
  \item \textit{The utility of any player depends solely on the members of the coalition to which the player belongs.}
  \item \textit{The coalitions form as a result of the preferences of the players over their possible coalitions’ set.} 
\end{enumerate}

\textbf{Proposition 1.} \textit{Our formulated problem matches the cooperative coalition formation game since it satisfies the conditions above.}
\begin{proof}
First, based on eq. (\ref{eq:10}), the utility function $U^{MML}_{j,i}$ depends solely on the MMLs in the coalition $\mathcal{C}_j$, $\forall~i \in \mathcal{C}_j$, $\forall~j \leq M$. Second, it is clear that a coalition is subject to modification if and only if an MML $i,~\forall~i \in \mathcal{C}_j$, would like to join a new coalition $\mathcal{C}_{j'}$ ,$\forall~j'\neq j$, in order to gain higher incentives based on its data similarity to the head $h_{j'}$, and its contribution in terms of data, computation, and reputation within the new coalition. Before examining how the cooperative game is applied to our proposed model, we introduce several definitions \cite{coalitionFormation1}.
\end{proof}

\textbf{Definition 1.} \textit{A coalition of metaverse meta-learners is denoted as $\mathcal{C}_j \subseteq \mathcal{N}^a$, where $j$ is the index of the coalition head.}

\textbf{Definition 2.} \textit{The set $\Pi=\{\mathcal{C}_1,...,\mathcal{C}_j,...,\mathcal{C}_M\}$ is a coalitional partition that includes all the MMLs in $\mathcal{N}^a$, where all $M$ coalitions are disjoint, i.e., $\mathcal{C}_j \cap \mathcal{C}_{j'}=\varnothing~\forall~j'\neq j$} and $\bigcup_{j=1}^{M} \mathcal{C}_j=\mathcal{N}^a$.

With the existence of $M$ heads, each MML has $M$ options to join any coalition and support the FML training that increases its utility. If all MMLs jointly contribute to the FML training of a single head, a grand coalition is created. A singleton coalition contains just one FL worker responsible for the FL training. When no MML volunteers for the FML training of a specific coalition, this coalition is represented by an empty set. It is worth mentioning that an MML does not participate in a coalition where its utility is negative.

In the cooperative game setting, each player is required to build preferences over all possible coalitions by calculating and ordering their utilities from each coalition. To evaluate these preferences, we define the concept of order relation as follows:

\textbf{Definition 3.} \textit{An order or preference relation $\succeq_i$ of any MML $i \in \mathcal{N}^a$ is defined as a reflexive, complete, and transitive binary relation over all coalition that it can join}

The above definition serves to evaluate the preference of a player $i \in \mathcal{N}^a$ over diverse coalitions in which it can join. Particularly,  $\mathcal{C}_j\succeq_i\mathcal{C}_{j'}$ indicates that the player $i$ prefers to join a coalition $\mathcal{C}_j \subseteq \mathcal{N}^a$ over a coalition $\mathcal{C}_{j'}\subseteq \mathcal{N}^a$ or at least has the same preference. Moreover, $\mathcal{C}_j\succ_i\mathcal{C}_{j'}$ indicates that the MML $i$ strictly prefers to join $\mathcal{C}_j$ over $\mathcal{C}_{j'}$. Accordingly, we introduce the following operation: 
 \begin{equation}
 \begin{split}
	\label{eq:12}
\mathcal{C}_j\succ_i\mathcal{C}_{j'} \iff U^{MML}_{j,i} > U^{MML}_{j',i}
 \end{split}
\end{equation}
We note that the complexity of the problem can be reduced by saving the history $h(i)$ of the coalitions that MML $i$ has joined before the formation of the current partition $\Pi$ using blockchain miners, and avoiding revisiting them in the future \cite{coalitionFormation1}. The preference of a coalition over another is task-specific, and it depends on many parameters, particularly the incentives that an MML can receive, its contribution in different coalitions in terms of data and computation, its similarity to the head, and the weight that the MML can give to other players in terms of overall reputation.

Given the preference relation $\succ_i$ of every meta-learner in the set of metaverse users $\mathcal{N}^a$, the cooperative coalition formation game can be formally defined as:

\textbf{Definition 4.} \textit{A hedonic or cooperative game is a coalitional game that is defined by a pair $(\mathcal{N}^a,\succ)$, where $\succ=\{\succ_1,...,\succ_i,...,\succ_{N^a}\}$} is the preference profile of different MMLs.

\textbf{Definition 5. (switch rule)} \textit{Given an initial partition $\Pi=\{\mathcal{C}_1,...,\mathcal{C}_j,...,\mathcal{C}_M\}$, an active MML $i \in \mathcal{C}_j$ chooses to leave its current coalition $\mathcal{C}_j$ and join another coalition $\mathcal{C}_{j'} \in \Pi,~j\neq j'$}, if and only if $\mathcal{C}_{j'}\cup \{i\}\succ \mathcal{C}_{j}$. Accordingly, the coalitions $\{\mathcal{C}_{j},\mathcal{C}_{j'}\}$ become $\{\mathcal{C}_{j} \setminus \{i\},\mathcal{C}_{j'}\cup \{i\}\}$.

For a given partition $\Pi$, the switch rule within a cooperative coalition formation game offers a mechanism that allows any active MML $i$ to leave its coalition $\mathcal{C}_{j} \in \Pi$ and align themselves with an alternative coalition $\mathcal{C}_{j'}\cup~\{i\}$, where $j'\neq j$. This transition occurs when the new coalition $\mathcal{C}_{j'}$ is strictly preferred over the current coalition $\mathcal{C}_{j}$ according to the established preference relation. As a result, the partition $\Pi$ becomes $\Pi'=\Pi\setminus\{\mathcal{C}_{j},\mathcal{C}_{j'}\}\cup\{\mathcal{C}_{j}\setminus\{i\},\mathcal{C}_{j'}\cup\{i\}\}$. The switch rule can be seen as a selfish decision from the active MML to shift to a new coalition, regardless of the consequences on the other players.

The coalition formation may be updated at each global round $k$. Particularly, we define an initial partition at $k=0$ equal to $\Pi^0=\{\mathcal{C}^0_1,...,\mathcal{C}^0_j,...,\mathcal{C}^0_M\}$. At the beginning of each round $k$, the partition may change to $\Pi^k=\{\mathcal{C}^k_1,...,\mathcal{C}^k_j,...,\mathcal{C}^k_M\}$ based on the switch rule. In other words, if the players' reputation or contribution to the training changes, the coalition formation potentially changes. $\Pi^{k+1}=\Pi^{k}$ means that no switch operation has been done, and any change of the users' behavior during the previous round did not impact the coalitional partition.

\textbf{Definition 6. (Nash-stability)} \textit{A partition $\Pi=\{\mathcal{C}_1,...,\mathcal{C}_M\}$ is Nash-stable, if $\mathcal{C}_j\succeq_i\mathcal{C}_{j'}\cup\{i\}$ where $j\neq j'$ and $\forall i \in \mathcal{C}_j$.}

$\Pi$ is Nash-stable if no player gains higher incentives from leaving its current coalition and joining another one or acting individually. This leads us to the concept of individual stability, defined as follows:

\textbf{Definition 7. (Individual-stability)} \textit{A partition $\Pi=\{\mathcal{C}_1,...,\mathcal{C}_M\}$ is Individually-stable, if there is no player $i \in \mathcal{C}_j$} where $\mathcal{C}_j \in \Pi$, such that $\mathcal{C}_{j'}\cup\{i\}\succeq_i\mathcal{C}_{j}$ and $\mathcal{C}_{j'}\succeq_k\mathcal{C}_{j'}\cup\{i\}$, $\forall k\in \mathcal{C}_{j'},~j\neq j'$.

An Individual-stability implies that no coalition $\mathcal{C}_j \in \Pi$ can be extended to $\mathcal{C}_{j}\cup\{i\}$, with a higher utility to the MML $i$ and without hurting the other players.
\subsection{Cooperative coalition formation algorithm} 
To reach the game's Nash-stability, we introduce the cooperative game formation algorithm, allowing the MMLs to select coalitions that maximize their utilities in the final partition. The process for cooperative coalition formation in the metaverse for meta-learners is outlined in Algorithm \ref{algo1}.
\begin{algorithm}[h]
\caption{Cooperative coalition formation algorithm}
\begin{algorithmic}[1]
\State \textbf{Input:} Set of active meta-learners $\mathcal{N}^a=\{1,..,i,...,N^a\}$
\State Set of heads $\mathcal{H}^{k-1}=\{h^{k-1}_1,...,h^{k-1}_j,...,h^{k-1}_M\}$
\State $\Pi^{k-1}=\{\mathcal{C}^{k-1}_1,...,\mathcal{C}^{k-1}_j,...,\mathcal{C}^{k-1}_M\}$
\State $\mathcal{H}^{k}=\emptyset$, $Nash-stability=0$
\State \textbf{Output:}  Final partition $\Pi^k=\{\mathcal{C}^k_1,...,\mathcal{C}^k_j,...,\mathcal{C}^k_M\}$
\State Initialize the final partition such that $\Pi^k = \Pi^{k-1}$
\State \textit{\underline{\textbf{Head selection:}}}
\For {$j=1:M$}
\State $\overline{T^j}=max(T^{i}_{max},~\forall i \in \mathcal{C}_j)$
\State $\underline{T^j}=min(T^{i}_{max},~\forall i \in \mathcal{C}_j)$
\State $h^{k}_j=max(\mathcal{R}^{global}_{j,i}\times(1-\frac{T^i_{max}-~\underline{T^j}}{\overline{T^j}-~\underline{T^j}}),~\forall i\in\mathcal{C}_j)$
\State $\mathcal{H}^{k}\longleftarrow \mathcal{H}^{k}\cup \{h^{k}_j\}$
\EndFor
\State \textit{\underline{\textbf{Switch rule:}}}
\While{$Nash-stability=0$}
\State $Nash-stability=1$
\For{\textbf{each }MML $i \in \mathcal{C}_j$ } 
\State Compute $U^{MML}_{j,i}$ by solving problem (\ref{eq:10})
\State and using $h^k_j \in \mathcal{H}^k$
\For{\textbf{each }Coalition $\mathcal{C}_{j'} \in \Pi^{k-1}, \forall j\neq j'$ }
\State compute $U^{MML}_{j',i}$ using $h^k_{j'}$ and $\mathcal{C}_{j'} \cup \{i\}$
\State compare $U^{MML}_{j,i}$ and $U^{MML}_{j',i}$
\If{$U^{MML}_{j',i} > U^{MML}_{j,i}$}
\State  MML $i$ leaves its current coalition:
\State $\mathcal{C}_j\longleftarrow \mathcal{C}_j \backslash \{i\}$
\State MML $i$ joins a new coalition:
\State $\mathcal{C}_{j'}\longleftarrow \mathcal{C}_{j'} \cup \{i\}$
\State Update the current partition $\Pi^{k}$:
\State $\Pi^{k}\longleftarrow \Pi^{k}\backslash \{\mathcal{C}_{j},\mathcal{C}_{j'}\}\cup\{\mathcal{C}_j \backslash \{i\},\mathcal{C}_{j'} \cup \{i\}\}$
\State $Nash-stability=0$
\EndIf
\EndFor
\EndFor
\EndWhile
\State \textbf{Return} Nash-Stable partition $\Pi^k=\{\mathcal{C}^k_1,...,\mathcal{C}^k_j,...,\mathcal{C}^k_M\}$ 
\end{algorithmic}
\label{algo1}	
\end{algorithm}

For all $k$, The partition $\Pi^k$ is first initialized to the previous partition $\Pi^{k-1}$ (line 7). Then, for each coalition $\mathcal{C}_j$, a new head is selected based on two criteria: the highest reputation and the lowest requirement in terms of task completion time (line 8 - line 14). Specifically, it is imperative that the heads are trusted to prevent any potential misleading of the training process. This concern arises if a head misbehaves and modifies its embedding, consequently misguiding data similarity to other players. Such behavior could ultimately affect the formation of coalitions. Furthermore, each head should be one of the players presenting a strict constraint in task completion time, so that the algorithm can guarantee to deliver the trained models within the time requirements of most metaverse users. Note that our algorithm will not ensure that all MMLs receive their models within their time requirements; instead, it ensures a trade-off between reputation and completion time (line 12).

The rest of the algorithm follows the switch rule outlined in Definition 5. Any MML $i \in \mathcal{C}_j$ can perform a switch operation if and only if it gets higher utility by joining another coalition $\mathcal{C}_{j'}$, where $j\neq j'$. More specifically, based on the updated reputations, the current contribution in terms of data and computation, and the embedding presented by the new head, each MML $i$ computes its utility from its current coalition using eq. (\ref{eq:10}) (line 19). Moreover, the MML $i$ evaluates its incentives from other coalitions that it can possibly join and rank them according to Definition 3 in order to determine its preference profile (line 21-line 23). If the MML achieves higher gains in another coalition, it leaves its current group and switches to the new one (line 24-line 28). Given any change in the partition, the coalitions are updated, and the Nash-stability is not yet reached (line 29-line 31). This process repeats until no MML has an incentive to leave its current coalition and join another one. At the end, the Nash-stable partition $\Pi^k$ that maximizes the utilities of all MMLs is returned (line 36). 

\textbf{Proposition 2.} \textit{Given any partition $\Pi^k=\{\mathcal{C}^k_1,...,\mathcal{C}^k_j,...,\mathcal{C}^k_M\}$ at a round $k$, the proposed cooperative game formation algorithm always converges to a nash-stable partition $\Pi^{k+1}$ composed of disjoint coalitions.}

\begin{proof}
The proposition is proved by contradiction. Suppose $\Pi^{k+1}$ is not Nash-stable. As per Definition 5, this implies the presence of switch operations contributing to increasing the utility of at least one MML by moving it from its present coalition to another. This contradicts the result of the proposed algorithm, where no other switch operation can increase the utility of any user. Hence, any partition $\Pi^{k+1}$ resulting from our cooperative game formation algorithm is final and Nash-stable and consequently individually stable.
\end{proof}
\section{Stackelberg game-based incentive mechanism for metaverse}\label{section:Stackelberg}
The interaction between the MSP and the MMLs is formulated as a single-leader multi-followers Stackelberg game. Within this framework, the MSP assumes the role of the leader, strategically determining the optimal competition incentives to encourage MMLs to execute meta-learning tasks. On the other hand, the MMLs, as followers, adapt their computation frequencies to increase their individual profits. The strategies of both the MSP and the MMLs are formulated as follows:
\begin{itemize}
    \item The MML's computing capacities in Stage II: Based on the reward strategy, MML $i \in \mathcal{C}_j$ determines its computing resources that maximize its profit which is given as: \newline
    \begin{equation}
    \begin{split}\raisetag{\baselineskip}
    U^{MML}_{j,i}(\delta_{j,i},I^j_{comp})=(S_{j,i}+(\frac{\delta_{j,i}-\underline{\delta_j}}{\overline{\delta_j}-\underline{\delta_j}})\times(\frac{|\mathcal{D}_{j,i}|}{\underset{k\in \mathcal{C}_j}{\sum |\mathcal{D}_{j,k}|}}))\\\times I^j_{comp} + \frac{\underset{k\in \mathcal{C}_j}{\sum H(\mathcal{R}_{j,k}))}}{|\mathcal{C}_j|} \times I_{rep} - c^{comp}_{j,i} - c^{comm}_{j,i}\qquad
    \end{split}
\end{equation}
For each coalition $\mathcal{C}_j$, the set of MMLs computing frequencies is $\delta_{j}=\{\delta_{j,1},\delta_{j,2},...,\delta_{j,|\mathcal{C}_j|}\}$, which serve to calculate the computation delays using eq. (\ref{eq:7}).
Accordingly, the round $k$ latency of the coalition $\mathcal{C}_j$ is:
\begin{equation}
	\label{eq:13}
T^{round}_{k,j}=max(T^{comp}_{j,i}+T^{comm}_{j,i},~\forall i \in \mathcal{C}_j)
\end{equation}
Each MML subgame problem is expressed as follows:

\textbf{Problem 1} (MML $i \in \mathcal{C}_j$ subgame):
\begin{equation}
	\label{eq:20}
  \begin{cases}
    \begin{array}{ll}
       \underset{\delta_{j,i}}\max&U^{MML}_{j,i}(\delta_{j,i},I^j_{comp})\\
       \mbox{subject to}& \underline{\delta^i} \leq \delta_{j,i} \leq \overline{\delta^i}\\
       &T^j_{max}\geq T^{comp}_{j,i}+T^{comm}_{j,i}\\
       &c^{comm}_{j,i} \ll c^{comp}_{j,i}\\
       &U^{MML}_{j,i}(\delta_{j,i},I^j_{comp})>0
    \end{array}
    \end{cases}
\end{equation}
where $\underline{\delta}$ and $ \overline{\delta}$ are the minimum and maximum computation capacities that can be offered by each MML. 
\item The MSP's rewarding strategy in Stage I: In Stage I, based on the resources offered by the MMLs, the MSP establishing an incentive strategy aimed at maximizing its utility, which is equal to:
\begin{equation}
\begin{split}\raisetag{\baselineskip}
U^{MSP}_j(I^j_{com}; \delta_{j})= \eta \sum_{i=1}^{|\mathcal{C}_j|} ln(T^j_{max}-T^{comp}_{j,i}-T^{comm}_{j,i})\\ \times H(\mathcal{R}_{j,i})-(S_{j,i}+(\frac{\delta_{j,i}-\underline{\delta_j}}{\overline{\delta_j}-\underline{\delta_j}})\times(\frac{|\mathcal{D}_{j,i}|}{\underset{k\in \mathcal{C}_j}{\sum |\mathcal{D}_{j,k}|}}))\times I^j_{comp} \\- \frac{\underset{k\in \mathcal{C}_j}{\sum H(\mathcal{R}_{j,k}))}}{|\mathcal{C}_j|} \times I_{rep},\qquad\qquad\qquad\qquad\qquad\qquad\qquad
\end{split}
\end{equation}
where $\eta$ is a profit weight parameter. The utility of the MSP depends on the computing resources the MMLs can offer, their reputations, and the incentives they receive. The $ln(.)$ reflects the MSP diminishing return on the completion time gain of each MML. The MSP subgame problem for coalition $\mathcal{C}_j$ is expressed as follows:

\textbf{Problem 2} (MSP subgame):
\begin{equation}
  \Bigg\{
    \begin{array}{ll}
       \underset{I^j_{com}}\max& U^{MSP}_j(I_{com}; \delta_{j})\\
       \mbox{subject to}& \underbar{$I_{com}$} \leq I^j_{com} \leq \overline{I_{com}} \\
    \end{array}
\end{equation}
where $\overline{I_{com}}$ and $\underbar{$I_{com}$}$ denote the maximum and minimum competition rewards, respectively.
\end{itemize}
We formulate the Stackelberg game by combining Problem 1 and Problem 2 with the objective of finding the Nash equilibrium solution.
\subsection{Game Equilibrium Analysis}
The Stackelberg equilibrium ensures the maximization of the MSP's utility within each coalition, given that the MMLs allocate their computational resources to the metaverse system following their optimal response strategy. In practical terms, this means that each MML's resource allocation strategy is designed to maximize its profit, taking into account the strategies of other players and the competition rewards offered by the MSP. The Stackelberg equilibrium in the coalition $\mathcal{C}_j$ can be expressed as follows:

\textbf{Definition 8.}
\textit{We denote $\delta_j^*=\{\delta_{j,1}^*,\delta_{j,2}^*,...,\delta_{j,|\mathcal{C}_j|}^*\}$ and $I_{com}^{j*}$ as the optimal computation capabilities of all the workers, and the optimal incentive offered by the MSP, respectively. The strategy $(\delta_j^*,I_{com}^{j*})$ is the Stackelberg equilibrium if:
\begin{equation}
\begin{split}
U^{MSP}_j(I^{j*}_{com}; \delta^*_{j}) \geq U^{MSP}_j(I^j_{com}; \delta^*_{j})\qquad\qquad \\
U^{MML}_{j,i}(\delta^*_{j,i},\delta^*_{j,-i},I^{j*}_{comp}) \geq U^{MML}_{j,i}(\delta_{j,i},\delta^*_{j,-i},I^{j*}_{comp}) 
\end{split}
\end{equation}}
where $\delta^*_{j,-i}$ is equal to $\delta^*_{j}\backslash\delta^*_{j,i}$. 
The backward induction is adopted to analyze the Stackelberg game.
\subsubsection{MMLs’ optimal strategies as equilibrium in Stage II}
Utilizing the incentive strategy $I^{j*}_{com}$ offered by the MSP to coalition $\mathcal{C}_j$, the workers in Stage II identify the optimal computation frequencies to maximize their profits.
The Lagrange function of (\ref{eq:20}) for each MML $i \in \mathcal{C}_j$ can be written as follows:

\textbf{Lagrangian equation:}
\begin{equation}
\begin{split}\raisetag{\baselineskip}
L(\delta_{j,i},\lambda_1,\lambda_2,\lambda_3)=(S_{j,i}+(\frac{\delta_{j,i}-\underline{\delta_j}}{\overline{\delta_j}-\underline{\delta_j}})\times(\frac{|\mathcal{D}_{j,i}|}{\underset{k\in \mathcal{C}_j}{\sum |\mathcal{D}_{j,k}|}}))\times I^j_{comp}\qquad\qquad \quad\\+ \frac{\underset{k\in \mathcal{C}_j}{\sum H(\mathcal{R}_{j,k}))}}{|\mathcal{C}_j|} \times I_{rep}-\rho \tau c_{j,i} |\mathcal{D}_{j,i}| \zeta (\delta_{j,i})^2+ \lambda_1(T^{comm}_{j,i} + \qquad\qquad\qquad\\\frac{\tau c_{j,i} |\mathcal{D}_{j,i}|}{\delta_{j,i}}-T^j_{max})+ \lambda_2 (\overline{\delta^i}-\delta_{j,i})+ \lambda_3(\delta_{j,i}-\underline{\delta^i})\qquad\qquad\qquad\qquad
\end{split}
\end{equation}
where $\lambda_1$, $\lambda_1$, and $\lambda_1$ are the Lagrangian multipliers. The optimal solution of $\delta_{j,i}$ is derived through the Karush-Kuhn-Tucker (KKT) conditions.

\textbf{KKT Conditions:}\\
\textit{1-Stationarity Conditions:}
By differentiating the Lagrangian with respect to $\delta_{j,i}$ and setting the derivative to zero, we can obtain:

\begin{equation}
\begin{split}\raisetag{\baselineskip}
\frac{\partial U^{MML}_{j,i}}{\partial \delta_{j,i}} = \frac{ \frac{|\mathcal{D}_{j,i}|}{\underset{k\in \mathcal{C}_j}{\sum |\mathcal{D}_{j,k}|}}
~I^j_{com}}{(\overline{\delta^j}-\underline{\delta_j})}  - 2\rho \tau c_{j,i} |\mathcal{D}_{j,i}| \zeta \delta_{j,i}
-\lambda_1 \frac{\tau c_{j,i} |\mathcal{D}_{j,i}|}{\delta_{j,i}^2} \\-\lambda_2 +\lambda_3=0\qquad\qquad\qquad\qquad\qquad\qquad\qquad\qquad\qquad\qquad
\end{split}
\end{equation}
\textit{2-Primal Feasibility:}
\begin{equation}
\begin{split}
T^j_{max}\geq  \frac{\tau c_{j,i} |\mathcal{D}_{j,i}|}{\delta_{j,i}}+T^{comm}_{j,i}\\
\overline{\delta^i}\geq \delta_{j,i}\\
\underline{\delta^i}\leq \delta_{j,i}
\end{split}
\end{equation}
\textit{3-Dual Feasibility:}
\begin{equation}
\begin{split}
\lambda_k \geq 0~ \forall k \in \{1,2,3\}\\
\end{split}
\end{equation}
\textit{4-Complementary Slackness:}
\begin{equation}
\begin{split}
\lambda_1(\frac{\tau c_{j,i} |\mathcal{D}_{j,i}|}{\delta_{j,i}}-T^j_{max})=0\\
\lambda_2 (\overline{\delta^i}-\delta_{j,i})=0\\
\lambda_3(\delta_{j,i}-\underline{\delta^i})=0
\end{split}
\end{equation}
~~\textbf{Possible cases:}\\
\textit{Case 1:~$T^j_{max} >  \frac{\tau c_{j,i} |\mathcal{D}_{j,i}|}{\delta_{j,i}}+T^{comm}_{j,i},~
\overline{\delta^i}> \delta_{j,i},~
\underline{\delta^i}< \delta_{j,i}$} \\
Here, all the constraints are strictly satisfied, implying $\lambda_1=\lambda_2=\lambda_3=0$.
\begin{equation}
\begin{split}
 \frac{ \frac{|\mathcal{D}_{j,i}|}{\underset{k\in \mathcal{C}_j}{\sum |\mathcal{D}_{j,k}|}}
~I^j_{com}}{(\overline{\delta_j}-\underline{\delta_j})}  - 2\rho \tau c_{j,i} |\mathcal{D}_{j,i}| \zeta \delta_{j,i} =0
\end{split}
\end{equation}
Then, the best response function of the MML $i \in \mathcal{C}_j$, i.e., $\delta_{j,i}$, can be derived following:
\begin{equation}
\begin{split}
\delta_{j,i}= \frac{\frac{|\mathcal{D}_{j,i}|}{\underset{k\in \mathcal{C}_j}{\sum |\mathcal{D}_{j,k}|}}
~I^j_{com}}{2\rho \tau c_{j,i} |\mathcal{D}_{j,i}| \zeta (\overline{\delta_j}-\underline{\delta_j})} 
\end{split}
\end{equation}
\textit{Case 2:~$T^j_{max}=  \frac{\tau c_{j,i} |\mathcal{D}_{j,i}|}{\delta_{j,i}}+T^{comm}_{j,i},~
\overline{\delta^i}\geq \delta_{j,i},~
\underline{\delta^i}\leq \delta_{j,i}$}\\
The first constraint is active, implying $\lambda_1$ can be non-zero. But, $\lambda_2=\lambda_3=0$
\begin{equation}
\begin{split}
\delta_{j,i}=\frac{\tau c_{j,i} |\mathcal{D}_{j,i}|}{T^j_{max}-T^{comm}_{j,i}}
\end{split}
\end{equation}
This case is feasible only if: 
\begin{equation}
\begin{split}
\lambda_1=\frac{\delta_{j,i}^2 \frac{|\mathcal{D}_{j,i}|}{\underset{k\in \mathcal{C}_j}{\sum |\mathcal{D}_{j,k}|}}
~I^j_{com}}{\tau c_{j,i} |\mathcal{D}_{j,i}|(\overline{\delta_j}-\underline{\delta_j})}  - 2\rho \zeta \delta^3_{j,i}>0
\end{split}
\end{equation}
\textit{Case 3:~$T^j_{max}>  \frac{\tau c_{j,i} |\mathcal{D}_{j,i}|}{\delta_{j,i}}+T^{comm}_{j,i},~
\overline{\delta^i}= \delta_{j,i}$}\\
The second constraint is active, implying $\lambda_2$ can be non-zero. But, $\lambda_1=\lambda_3=0$. This case is feasible, only if: 
\begin{equation}
\begin{split}
 \lambda_2 = \frac{ \frac{|\mathcal{D}_{j,i}|}{\underset{k\in \mathcal{C}_j}{\sum |\mathcal{D}_{j,k}|}}
~I^j_{com}}{(\overline{\delta_j}-\underline{\delta_j})}  - 2\rho \tau c_{j,i} |\mathcal{D}_{j,i}| \zeta \delta_{j,i}>0
\end{split}
\end{equation}
\textit{Case 4: ~$T^j_{max}>  \frac{\tau c_{j,i} |\mathcal{D}_{j,i}|}{\delta_{j,i}}+T^{comm}_{j,i},~
\underline{\delta^i}= \delta_{j,i}$}\\
The third constraint is active, implying $\lambda_3$ can be non-zero. But, $\lambda_1=\lambda_2=0$. This case is feasible, only if: 

\begin{equation}
\begin{split}
\lambda_3=2\rho \tau c_{j,i} |\mathcal{D}_{j,i}| \zeta \delta_{j,i}-\frac{ \frac{|\mathcal{D}_{j,i}|}{\underset{k\in \mathcal{C}_j}{\sum |\mathcal{D}_{j,k}|}}
~I_{com}}{(\overline{\delta_j}-\underline{\delta_j})}>0
\end{split}
\end{equation}
\textit{Case 5:~$T^j_{max}=  \frac{\tau c_{j,i} |\mathcal{D}_{j,i}|}{\delta_{j,i}}+T^{comm}_{j,i},~
\overline{\delta^i}= \delta_{j,i}$}\\
The first and second constraints are active, implying $\lambda_1$ and $\lambda_2$ can be non-zero. But, $\lambda_3=0$. This case is feasible, only if: 
\begin{equation}
\begin{split}
\overline{\delta^i}=\frac{\tau c_{j,i} |\mathcal{D}_{j,i}|}{T^j_{max}-T^{comm}_{j,i}}\qquad\qquad\qquad\\
\text{and}\qquad\qquad\qquad\qquad\qquad\qquad\qquad\qquad\qquad\\
\frac{ \frac{|\mathcal{D}_{j,i}|}{\underset{k\in \mathcal{C}_j}{\sum |\mathcal{D}_{j,k}|}}
~I^j_{com}}{(\overline{\delta_j}-\underline{\delta_j})}  - 2\rho \tau c_{j,i} |\mathcal{D}_{j,i}| \zeta \delta_{j,i}>0
\end{split}
\end{equation}
\textit{Case 6:  ~$T^j_{max}=  \frac{\tau c_{j,i} |\mathcal{D}_{j,i}|}{\delta_{j,i}}+T^{comm}_{j,i},~
\underline{\delta^i}= \delta_{j,i}$}\\
The first and third constraints are active, implying $\lambda_1$ and $\lambda_3$ can be non-zero. But, $\lambda_2=0$. This case is feasible, only if: 
\begin{equation}
\begin{split}
\underline{\delta^i}=\frac{\tau c_{j,i} |\mathcal{D}_{j,i}|}{T^j_{max}-T^{comm}_{j,i}}\qquad\qquad\qquad\\
\text{and}\qquad\qquad\qquad\qquad\qquad\qquad\qquad\qquad\qquad\\
\frac{\delta_{j,i}^2 \frac{|\mathcal{D}_{j,i}|}{\underset{k\in \mathcal{C}_j}{\sum |\mathcal{D}_{j,k}|}}
~I^j_{com}}{\tau c_{j,i} |\mathcal{D}_{j,i}|(\overline{\delta_j}-\underline{\delta_j})}  - 2\rho \zeta \delta^3_{j,i}>0
\end{split}
\end{equation}
\subsubsection{MSP optimal reward strategy in Stage I}
Building upon the optimal computation capabilities of MMLs in Stage II, the MSP aims at maximizing its utility in Stage I by acting as the leader.

\textbf{Proposition 3.} \textit{The uniqueness of the proposed Stackelberg game’s equilibrium can be guaranteed.}

\begin{proof}
From case 2 to case 6, the second-order derivative of the MSP’s utility function related to the MML $i \in \mathcal{C}_j$ is equal to 0. Regarding case 1, the first-order derivative of the MSP’s utility function related to each MML $i \in \mathcal{C}_j,$ is shown as follows:
\begin{equation}
\label{a}
\begin{split}
\frac{\partial U^{MSP}_j}{\partial I^j_{com} } = \eta H(\mathcal{R}_{j,i}) \times \qquad\qquad\qquad\qquad\qquad\quad\\
\frac{\tau c_{j,i} |\mathcal{D}_{j,i}|}{I^{j2}_{com} E_{j,i}(1+T^j_{max}-T^{comm}_{j,i})-\tau c_{j,i} |\mathcal{D}_{j,i}|I^j_{com}}\\-2 \frac{I^j_{com} E_{j,i}}{(\overline{\delta_j}-\underline{\delta_j})} \frac{|\mathcal{D}_{j,i}|}{\underset{k\in \mathcal{C}_j}{\sum |\mathcal{D}_{j,k}|}}+ \frac{\underline{\delta^j}}{(\overline{\delta_j}-\underline{\delta_j})} \frac{|\mathcal{D}_{j,i}|}{\underset{k\in \mathcal{C}_j}{\sum |\mathcal{D}_{j,k}|}}-S_{j,i}
\end{split}
\end{equation}
with  $E_{j,i}= \frac{\frac{|\mathcal{D}_{j,i}|}{\underset{k\in \mathcal{C}_j}{\sum |\mathcal{D}_{j,k}|}}
~I^j_{com}}{2\rho \tau c_{j,i} |\mathcal{D}_{j,i}| \zeta (\overline{\delta_j}-\underline{\delta_j})}$ \\
The second-order derivative of the MSP’s utility function is:

\begin{equation}
\label{b}
\begin{split}
\frac{\partial^2 U^{MSP}_j}{\partial I^{j2}_{com} } =-\eta H(\mathcal{R}_{j,i})\times \qquad\qquad\qquad\qquad\qquad\quad\\\frac{2 \tau c_{j,i} |\mathcal{D}_{j,i}|E_{j,i} (1+T^j_{max}-T^{comm}_{j,i}) + (\tau c_{j,i} |\mathcal{D}_{j,i}|)^2}{(E_{j,i} (1+T^j_{max}-T^{comm}_{j,i})I^2_{comp}-\tau c_{j,i} |\mathcal{D}_{j,i}| I^2_{com})^2} \\ -2 \frac{E_{j,i}}{(\overline{\delta_j}-\underline{\delta_j})} \frac{|\mathcal{D}_{j,i}|}{\underset{k\in \mathcal{C}_j}{\sum |\mathcal{D}_{j,k}|}}\qquad\qquad\qquad\qquad\qquad\qquad\quad
\end{split}
\end{equation}
The second-order derivative of the MSP’s utility function related to the MML $i \in \mathcal{C}_j$ is always negative since $T^j_{max}>T^{comm}_{j,i}$, which indicates that $U^{MSP}_j$ is a concave function. Hence, for each coalition $\mathcal{C}_j,~\forall~\mathcal{C}_j \in \Pi$, the MSP has a unique optimal solution, which can be effectively determined using the bisection method \cite{bisection}. From the MSP's optimal strategy, the MMLs' optimal strategies can be derived. As a result, the proposed model ensures the achievement of the Stackelberg equilibrium.
This equilibrium allows the MSP to maximize its utility and the MMLs to guarantee optimal profit, without incentive to modify their strategies for increased gains.
\end{proof}
\subsection{Dual Game-theoretic Federated Meta-Learning algorithm}
\begin{algorithm}[h]
\caption{Game-theoretic Federated Meta-learning (GFML) algorithm for Metaverse}
\begin{algorithmic}[1]
\State \textbf{Input:} $\mathcal{N},~M,~\alpha,~\beta,~\tau,~\rho,~\lambda,~\phi,~\chi,~\gamma,~\mathcal{R}_{th},~I_{rep},$
\State $\overline{I_{comp}},~\underline{I_{comp}}.$
\State Select $\mathcal{H}^0=\{h^0_1,h^0_2,...,h^0_M\}$ randomly.
\State Create $\Pi^0=\{\mathcal{C}_1^0,\mathcal{C}_2^0,...,\mathcal{C}_M^0\}$ randomly.
\State MSP initializes a model $\theta^0$ and sends it to all heads.
\State Calculate embeddings $\omega^0_i~\forall~  i \in \mathcal{C}_j$ using the head $h^0_j$
\State  model.
\State $n=1$
\For { \text{round} $k=1:K-1$}
\State $\Pi^k$~=Run Algorithm \ref{algo1}
\If {$k ==n\times \chi$}
\State \text{Calculate embeddings $\omega^k_i~\forall~  i \in \mathcal{C}_j$} using 
\State the head $h^k_j$ model.
\EndIf
\State $n\longleftarrow n+1$
\For {coalition $j=1:M$} 
\State Find $I^j_{com}$ based on $\frac{\partial U^{MSP}_j}{\partial I^j_{com} }$$=0$
\For {MML $i=1:|\mathcal{C}_j|$}
\State Initialize: $\theta_j^{k,0}\longleftarrow\theta_j^{k-1}$ and $u^k_{j,i}\longleftarrow 0$
\State Find $\delta_{j,i}$ using KKT conditions.
\For {local step $t=1:\tau$}
\State Compute stochastic gradient $\Tilde{\nabla} F_{j,i}(\theta)$
\State using (\ref{eq:3}).
\State Update local model $\theta^{k,t+1}_{j,i}$ using (\ref{eq:2})
\State Update training contribution:
\State $u^k_{j,i}\longleftarrow u^k_{j,i}+ \parallel \Tilde{\nabla} F_{j,i}(\theta_{j,i}^{t,k})\parallel^2$
\State $-2(1+\frac{1}{\sqrt{|\mathcal{D}_{j,i}|}})\parallel \Tilde{\nabla} F_{j,i}(\theta_{j,i}^{t,k})\parallel$
\EndFor
\State Set $\theta_{j,i}^{k}\longleftarrow\theta_{j,i}^{k,\tau}$ and send to $h^k_j$
\State Send $u^k_{j,i}$ to blokchain and update global \State reputation based on contribution $\vartheta^k_{j,i}$ in (\ref{eq:contribution}) 
\EndFor
\State Compute the global model within $h^k_j$ using (\ref{eq:5})
\EndFor
\EndFor
\end{algorithmic}
\label{algo2}	
\end{algorithm}

Our framework employs dual game-theory, as summarized in Algorithm \ref{algo2}, named GFML. The algorithm begins with an initialization step, involving random head selection and coalition formation. Subsequently, during each training round, coalitions undergo adjustments through the cooperative coalition formation game, based on the embeddings generated from the heads models, reputations, and allocated resources. Once coalitions are established, we fine-tune the computations of MMLs  and the rewards provided by the MSP through the Stackelberg game strategy. This process aims to maximize the utility of the service provider and increase user profits. Each MML then carries out local training, guided by its computing frequency derived from the game. Local model results are aggregated within the heads, and reputations are updated based on their respective contributions. As we move forward to the next round, the coalition formation game takes into account these new reputations and computation capabilities, ensuring a dynamic and adaptive approach to our framework's operation.
\section{Performance evaluation}\label{section:evaluation}
\begin{figure*}[h]
	\centering
	\mbox{
	 \subfigure[\label{accuracy_after}]{\includegraphics[scale=0.44]{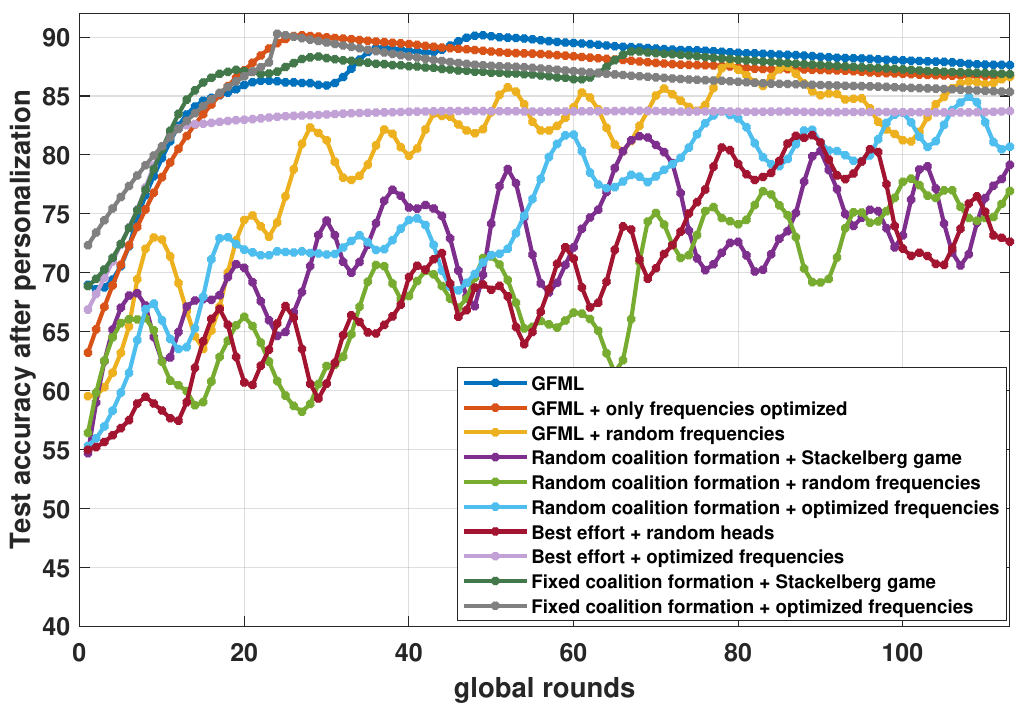}}
	 \subfigure[\label{accuracy_before}]{\includegraphics[scale=0.44]{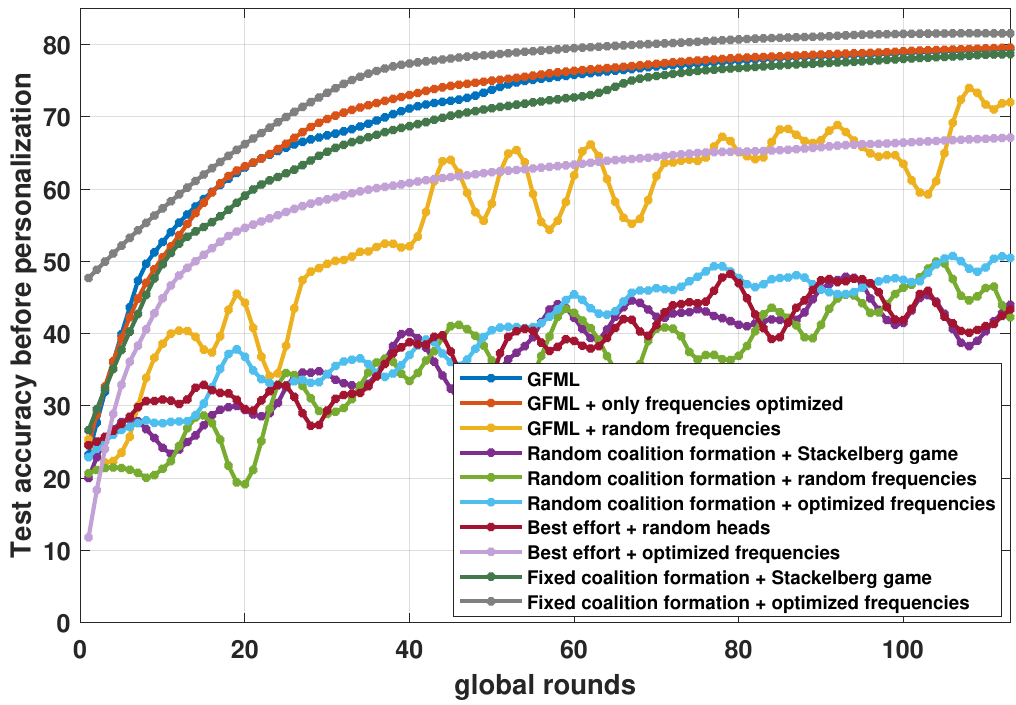}}}\\
  \vspace{-7pt}
		\mbox{\subfigure[\label{loss}]{\includegraphics[scale=0.44]{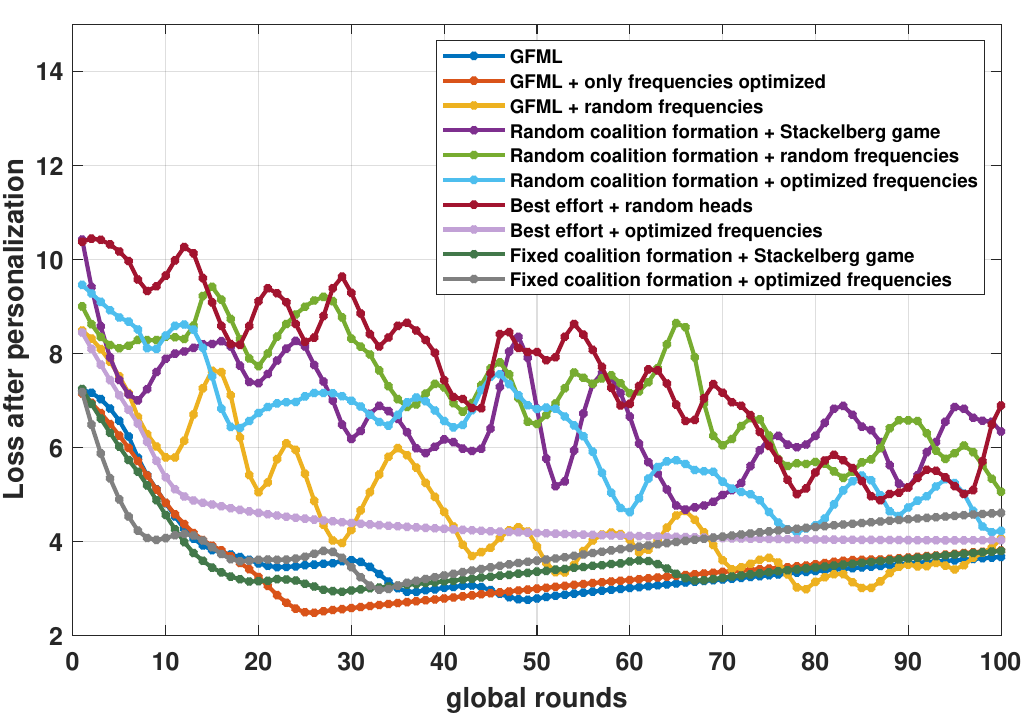}}}
	\caption{Performance of our proposed framework compared to existing baselines.}
	\label{performance}
\end{figure*}
This section presents the numerical results of the cooperative coalition formation game and  Stackelberg game-based incentive mechanism for reliable meta-learning in the metaverse. The parameters values are given in Table \ref{parameters}.

\begin{table}[t]
	\caption{Parameter settings.}
	\label{parameters}
 \centering
 \footnotesize
	\begin{tabular}{C{5cm}|L{1.2cm}}
		\hline   Parameters & values \\
        \hline   
        Number of MMLs $|\mathcal{N}|$ & 40  \\   
        Number of coalitions $M$  & 4 \\  
        Size of the model $W$ &  10 Mb  \\  
        Meta-learning rate $\alpha$ & $10^{-3}$  \\ 
        Step size $\beta$ & $10^{-2}$  \\ 
        Batch size & $40$  \\ 
         Number of iterations $\tau$ & 10 \\
       Unit computation cost $\rho$ & 0.1 \\
       Effective capacitance $\zeta$ & $10^{-27}$ \\
       CPU cycles/bit $c$ & 10   \\
       Max computation capacity $\overline{\delta^i}~\forall i\in\mathcal{N}$ & $10^9$ \\
       Min computation capacity $\underline{\delta^i}~\forall i\in\mathcal{N}$ & $10^7$  \\
       Mean transmission rate $B$ & $[5,10]Mhz$  \\ 
       Tolerated completion time $T_{max}$ & $[11,20]$ s  \\ 
       Reputation declining factor $\lambda$ & 0.7  \\
       Reputation weighting parameter $\phi$ & 0.2  \\
       Competition incentive $I_{comp}$ & [5, 15]  \\
       Reputation incentive $I_{rep}$ & 5  \\
       Default reputation utility $\gamma$ & 0.5  \\
       Reputation threshold $\mathcal{R}_{th}$& 0.5  \\
       MSP weighting parameter $\eta$& 25  \\
		\hline 
	\end{tabular}
\end{table}
In our evaluation, we use two widely used benchmarks: Fashion-MNIST and MNIST. The data from these benchmarks are distributed across 40 devices, with each device holding samples from two randomly chosen classes and the number of samples per class adhering to a quantity-based label distribution. We utilize the FedLab library\footnote{https://fedlab.readthedocs.io/en/master/index.html} for data partitioning. Randomly selected 90\% of the devices serve as active MMLs, leaving the remainder passive. For each device, the local dataset is split, allocating 90\% for training and 10\% for testing. Furthermore, our starting model is a deep neural network with Exponential Linear Unit (ELU) activation function, containing three fully connected layers with sizes 80, 60, and 10. It is worth noting that employing more complex neural networks would improve the results of our framework and baseline methods, yet relative performance would stay the same. For aggregation, we use the Personalized FedAvg (Per-FedAvg HF) algorithm proposed in \cite{FML}. To validate the performance of our proposed framework, we consider multiple baselines:
\begin{itemize}
    \item \textit{GFML + optimized frequencies:} Our GFML framework is deployed without the Stackelberg game. Instead, all competition incentives $I^j_{com}~\forall~ \mathcal{C}_j \in \Pi$ are fixed to 15, and the frequencies of users are optimized with the objective of maximizing their payoff $U^{MML}_{j,i}$ following the problem (\ref{eq:20}). It is important to note that the frequencies are optimized without considering the utility of the MSP, and users with the lowest $T_{max}$ are selected as heads.
    \item \textit{GFML + Random frequencies:} Our GFML framework is deployed without Stackelberg game. Instead, the competition incentives $I^j_{com}$ are fixed to 15 and the heads and frequencies are selected randomly. We note that users are discarded from the training if they cannot respect the completion time threshold of any coalition.
    \item \textit{Random coalition formation + Random frequencies:} Coalitions, frequencies, and heads are selected randomly \cite{FML,meta-learning2}.
    \item \textit{Random coalition formation + Stackelberg game:} Coalitions and heads are selected randomly, but the resources and incentives are managed using Stackelberg game.
    \item \textit{Random coalition formation + Optimized frequencies:} Coalitions and heads are selected randomly, but incentives $I^j_{com}$ are fixed to 15 and resources are optimized following the problem in equation (\ref{eq:20}).
    \item \textit{Best effort + Random head:} All users' models are aggregated together in each round and the head is selected randomly 
    \item \textit{Best effort + Stackelberg game:} The user with lowest $T_{max}$ is selected as aggregator. Incentives and resources are managed using Stackelberg game.
    \item \textit{Best effort + optimized frequencies:} The user with lowest $T_{max}$ is selected as aggregator. Incentives are fixed and resources optimized without considering MSP utility.
    \item \textit{Fixed coalitions + Stackelberg game:} This baseline is a non-uniform strategy, where heads with lowest $T_{max}$ are selected, and users are clustered based only on the proposed embeddings similarities. Stackelberg game is used to manage incentives and resources.
    \item \textit{Fixed coalitions + optimized frequencies:} Compared to the previous baseline, the incentives are fixed and only frequencies are optimized.
\end{itemize}

\begin{figure*}[h]
\centering
	\includegraphics[scale=0.48]{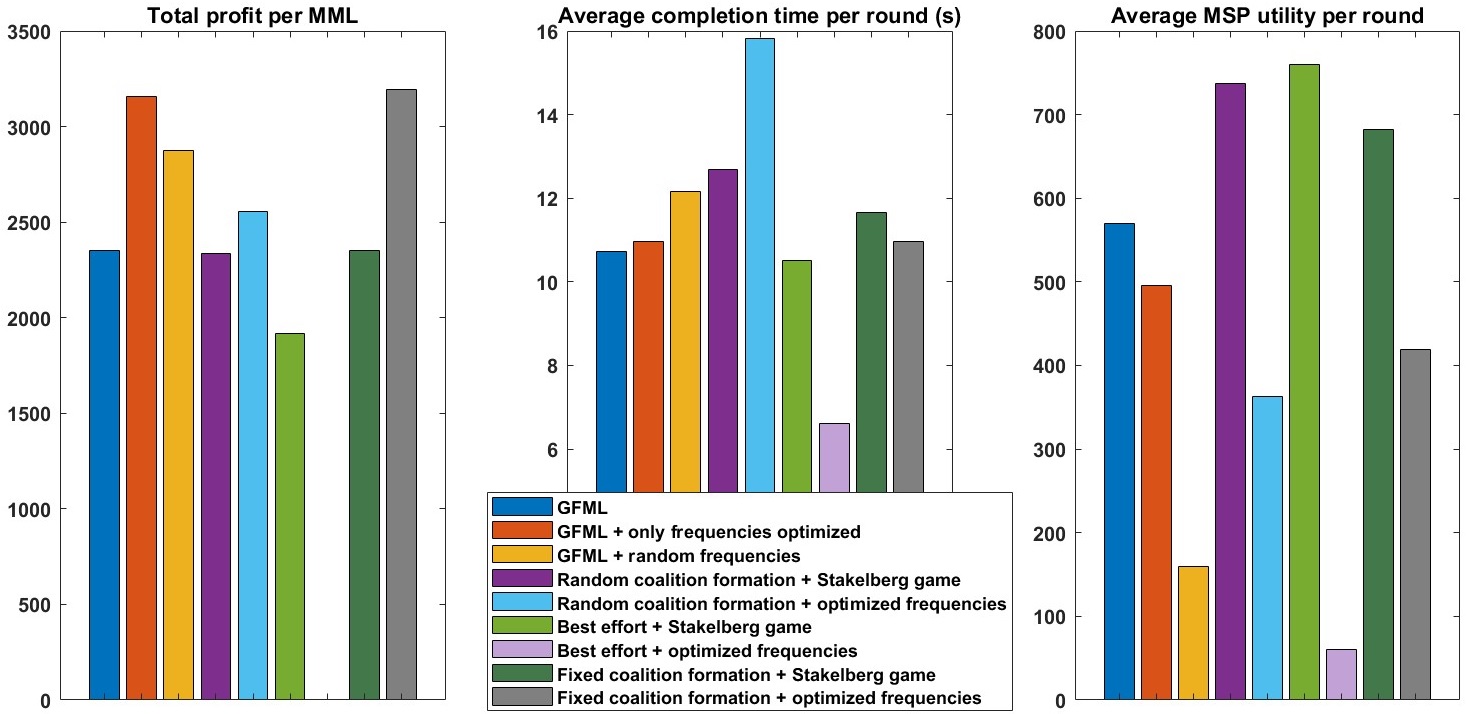}
	\caption{MSP and users utilities in our framework compared to existing works.}
	\label{utilities}
\end{figure*}

\begin{figure}[h]
\centering
	\includegraphics[scale=0.45]{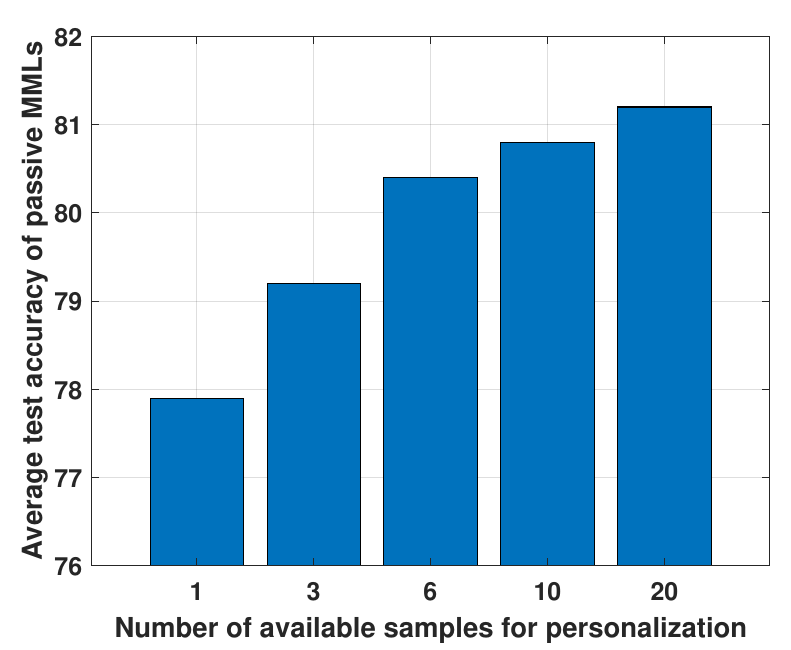}
	\caption{Test accuracy of passive MMLS.}
	\label{passive}
\end{figure}

\begin{figure*}[h]
	\centering
	\mbox{
	 \subfigure[\label{accuracy}]{\includegraphics[scale=0.4]{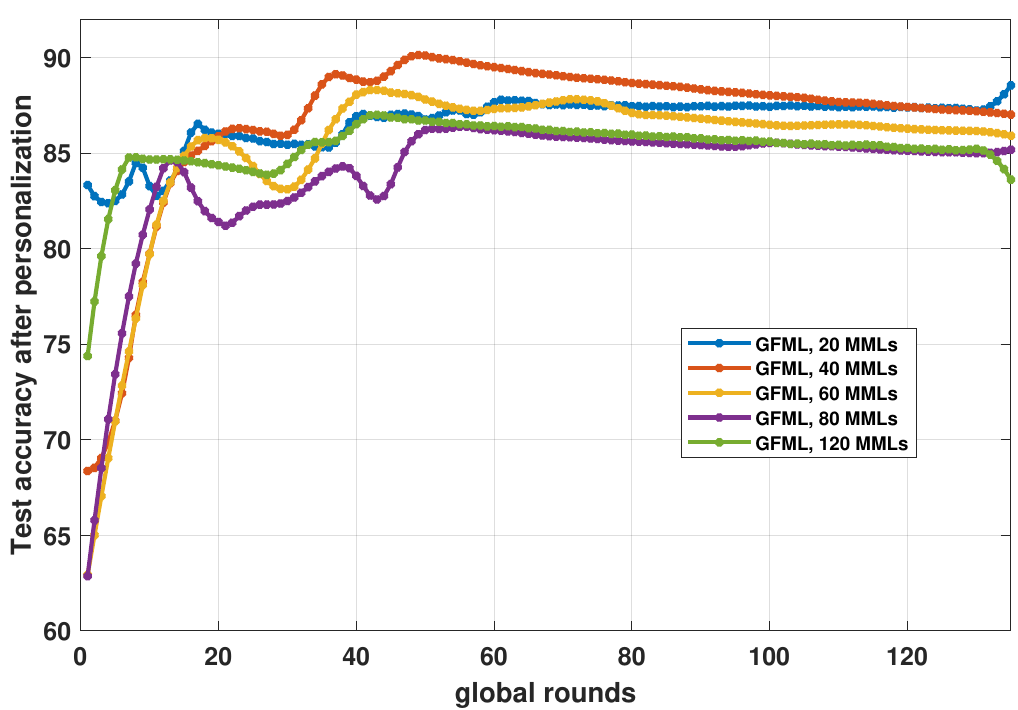}}
  \subfigure[\label{Utilities_NumberUsers}]{\includegraphics[scale=0.4]{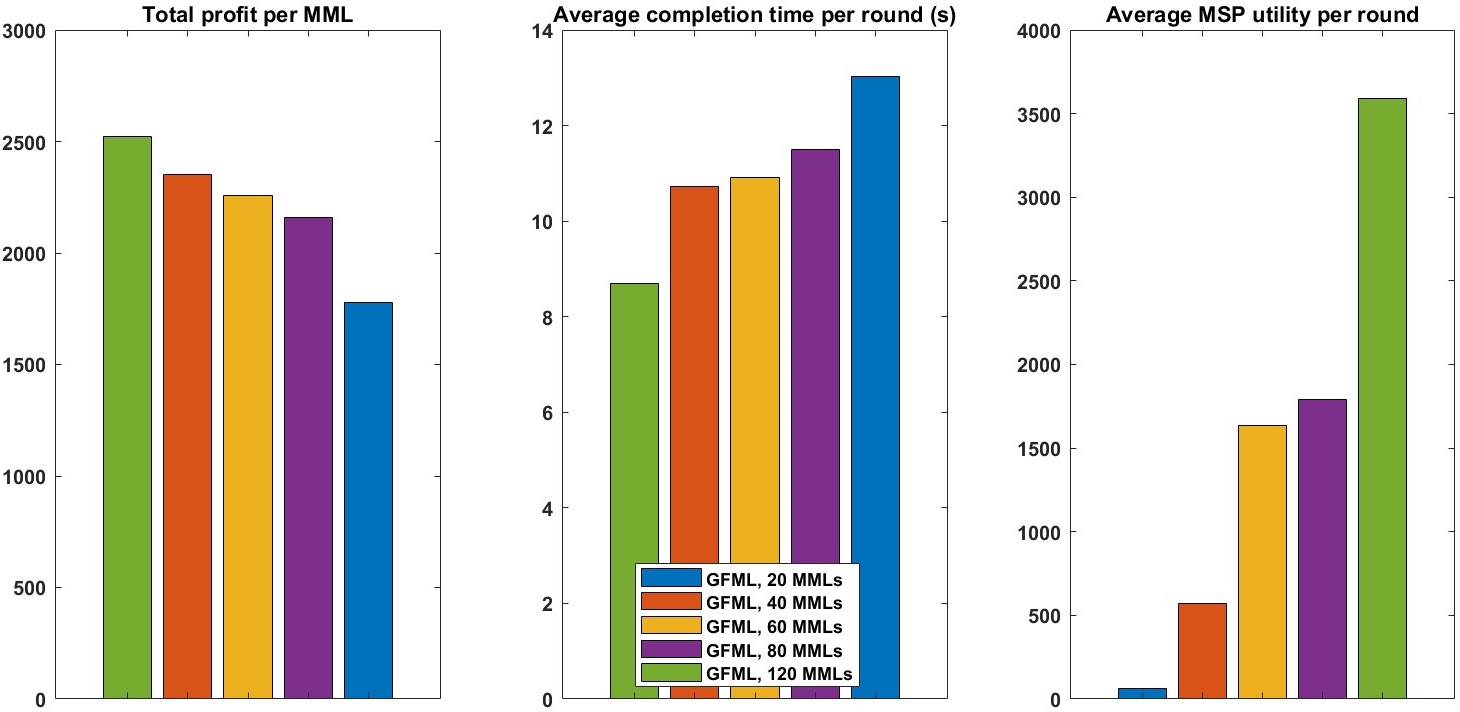}}}
	\caption{Impact of the number of users on the performance of our proposed framework.}
	\label{NumberUsers}
\end{figure*}
\begin{table*}[]
\caption{MSP and users utilities in our framework compared to existing works (numerical presentation).}
\label{tab:res}
\resizebox{\textwidth}{!}{%
\footnotesize
\begin{tabu}{|l|l|l|l|}
\hline
Framework & Total profit per MML & Average completion time per round (s) & Average MSP utility per round \\ \hline
GFML & 2350 & 10.72 & 570 \\ \hline
GFML + only frequencies optimized & 3158.7 & 10.96 & 496 \\ \hline
GFML + random frequencies & 2876.24 & 12.15 & 159.86 \\ \hline
\begin{tabular}[c]{@{}l@{}}Random coalition formation + \\ Stackelberg game\end{tabular} & 2336.62 & 12.68 & 737.77 \\ \hline
\begin{tabular}[c]{@{}l@{}}Random coalition formation + \\ optimized frequencies\end{tabular} & 2554.2 & 15.82 & 363.35 \\ \hline
Best effort + Stackelberg game & 1915 & 10.5 & 760.2 \\ \hline
Best effort + optimized frequencies & 2.01 & 6.62 & 60 \\ \hline
\begin{tabular}[c]{@{}l@{}}Fixed coalition formation + \\ Stackelberg game\end{tabular} & 2350.31 & 11.67 & 682 \\ \hline
\begin{tabular}[c]{@{}l@{}}Fixed coalition formation + \\ optimized frequencies\end{tabular} & 3194.7 & 10.97 & 419.1 \\ \hline
\end{tabu}%
}
\end{table*}
\begin{figure}[h]
\centering
	\includegraphics[scale=0.4]{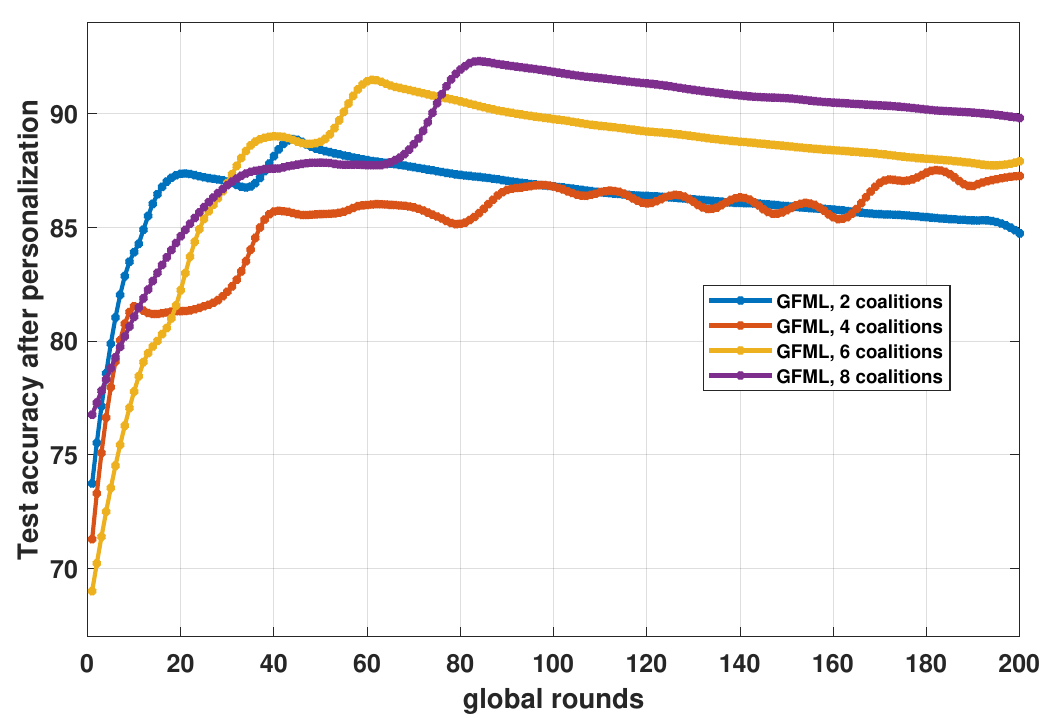}
	\caption{Impact of the number of coalitions on the performance of our proposed framework.}
	\label{NumberCoalitions}
\end{figure}
\subsection{Numerical analysis of the GFML performance under an honest environment}
Fig. \ref{accuracy_after} and Fig. \ref{loss} show the performance of the personalized models obtained by the active users in terms of average test accuracy and loss. We can see that our proposed framework outperforms all the other baselines and strategies. Particularly, the best effort strategy presents a stable and fast convergence when the frequencies are optimized and heads are not randomly selected; however, the achieved accuracy is low. This can be explained by the fact that all users with diverse tasks and high statistical heterogeneity are aggregated together, undermining the global training outcome. When the head is randomly selected, the constraint on completion time $T_{max}$ changes at each training round; and since resources are not optimized, users that do not respect the constraint are excluded, which makes the learning unstable. This is not the case in our framework, where coalitions are formed based on data statistical similarity and heads are wisely selected. 

We can also see that the random coalition formation is unstable and has a low convergence speed,  which is also due to bias and statistical heterogeneity. In GFML, when we opt for random frequency generation instead of optimized computational resources and incentives using Stackelberg game, our proposed scheme presents unstable and slower convergence and lower accuracy.  This is explained by the inconsistency in computational contributions, leading to reputation fluctuations and consequently impacting the coalition formation game. Furthermore, the MMLs that cannot respect the completion time constraint are discarded from the meta-learning process, depriving them of a more personalized experience and hence reducing the average training performance. The same behavior can also be seen for random clustering schemes with random frequencies compared to optimized computation resources.

When our GFML scheme is deployed with fixed incentives and optimized computation resources without considering the profit of the MSP, the system exhibits stable performance, high convergence speed, and satisfactory accuracy. However, using Stackelberg game presents a slightly better performance, as the objective of the game is not only to maximize the payoff of the users but also to increase the utility of the MSP. Particularly, the utility of the MSP is lower when $I^j_{com}~\forall~\mathcal{C}_j~ \in \Pi$ are higher. The objective of the Stackelberg game is to minimize $I^j_{com}$ while maintaining a satisfactory performance. As such, when $I^j_{com}$ are lower, the impact of the overall reputation weighted by $I_{rep}$ is maximized (see eq. \ref{eq:10}), which enhances the performance of the training. We remind that the reputation of MMLs provides insights into their training contributions. 

Finally, the fixed coalition formation demonstrates both stability and high performance  due to the unchanging clusters formed based on statistical similarity. This underscores the effectiveness of our proposed embeddings similarity method.  Yet, GFML relies on both embeddings similarity and task-specific reputation, offering insights into each user's potential contribution to the current task based on past task similarities. As a result, our framework excels in performance. This assertion is confirmed in Fig. \ref{accuracy_before} which depicts the average accuracy of global models prior to MML personalization. In this figure, the fixed coalition formation scheme achieves better results, as the trained models are less task-specific. Upon personalization, our proposed scheme surpasses it. We also observe that, after personalization, the fixed coalition formation using the Stackelberg game yields superior performance compared to optimizing solely the MMLs' payoffs for the same reasons mentioned earlier.

Fig. \ref{utilities} and its numerical presentation in Table \ref{tab:res} illustrate the average total payoff of MMLs, the average completion time per round, and the average MSP utility per round. We observe that strategies, including GFML and random and fixed coalition formation, which focus solely on maximizing the profit of MMLs, effectively achieve their goal of enhancing user payoffs owing to the optimized frequencies balancing the profit and energy consumption. However, the presented MSP utility is low, as its maximization is not considered in the resource allocation optimization or the coalition formation game.
On the other hand, strategies employing the Stackelberg game successfully strike a balance between MMLs' profit and the utility of the MSP. In particular, compared to GFML, random and fixed coalition formations yield slightly higher utility for the MSP due to users providing lower computation resources, resulting in lower costs for the service provider. However, due to these low frequency resources, their average completion time tends to be among the highest. This is not the case of our framework, constrained by stricter time constraints imposed by the dynamically selected heads. Such constraint is looser for the random and fixed coalitions where heads are either randomly selected or fixed, potentially requiring larger completion times and leading to lower computation capacities. 

The high task latency observed in the random formation schemes can be attributed to the arbitrary selection of heads, which might tolerate higher completion times. Our GFML approach with random frequencies results also in high completion times, primarily due to the potentially low computation resources generated arbitrary. While this approach may benefit users by lowering energy costs, it leads to minimal MSP utility due to training instability resulted from unstable coalition formation highly impacted by random frequencies. In the best effort scenario, since user utility depends partially on data size (see Eq. \ref{eq:10}) and given that all MMLs's models are aggregated together and individual user data might be negligible relative to the total, the positive profit might be minor. In this case, the energy costs could result in negative incentives. Hence, when only users' profits are optimized, the optimization boosts the computational frequency until achieving a near-zero utility, resulting in low task latency but incentives that deter active participation in metaverse services. Stackelberg game helps to achieve a balance between the MSP utility and the users' profits in best effort strategy, although the overall system welfare remains lower than in our proposed system  due to training inefficiency caused by statistical heterogeneity. To conclude, our framework presents the highest training performance while striking a balance between the profit of the users and the utility of MSP, in addition to ensuring a low average completion time. This owes to both coalition formation game relying on data similarity and historical reputation for similar tasks, and stackelberg game optimizing not only MMLs resources but also MSP profits.

Fig. \ref{passive} illustrates the performance of the trained models personalized by the passive MMLs that did not participate actively in the training. From the results, even with a single sample for personalization, users can attain high-performing models without direct training participation. This highlights the meta-learning efficacy in supporting immersive user experiences within the metaverse, 
through optimizing models for quick adaptability to unforeseen tasks and information, enabling  prompt edge learning and efficient personalization with minimal data input. Possessing additional samples for personalization further refines the model's accuracy, bringing its performance in line with active MMLs. 

Fig. \ref{accuracy} depicts the test accuracy of our system as we vary the number of users. Notably, as the number of users decreases, accuracy tends to increase. It is worth noting that we maintain a fixed number of coalitions at 4. Fewer users imply fewer tasks and reduced complexity in terms of heterogeneity. Therefore, dividing users into 4 coalitions proves effective in addressing data and distribution disparities. However, when dealing with a larger number of highly heterogeneous users, clustering them into only 4 coalitions fails to manage task differences effectively. In this case, a higher number of coalitions is recommended when serving a larger number of metaverse users. Fig. \ref{Utilities_NumberUsers} illustrates the profit of the MMLs, the utility of the MSP, and the average completion time, when varying the number of users. It is expected that with a larger user base, the utility of the MSP, calculated as the sum of utilities from each user, increases. Moreover, as the number of MMLs increases, the probability of encountering users with stricter completion time requirements also rises. Consequently, we observe that the average task completion time is the lowest when the system consists of 120 MMLs. Accordingly, their profits increase because they offer higher computation resources to meet strict time constraints.

Fig. \ref{NumberCoalitions} shows the impact of varying the number of coalitions on the performance of our framework, when fixing the number of users to 40. It is clear that when the number of coalitions is higher, the training performance is better. This can be attributed to the fact that clustering contributes to capture the differences between different tasks. Consequently, distributing users across more coalitions enhances task granularity and, consequently, training efficiency.

\begin{figure*}[h]
	\centering
	\mbox{
	 \subfigure[\label{misbehavior5}]{\includegraphics[scale=0.4]{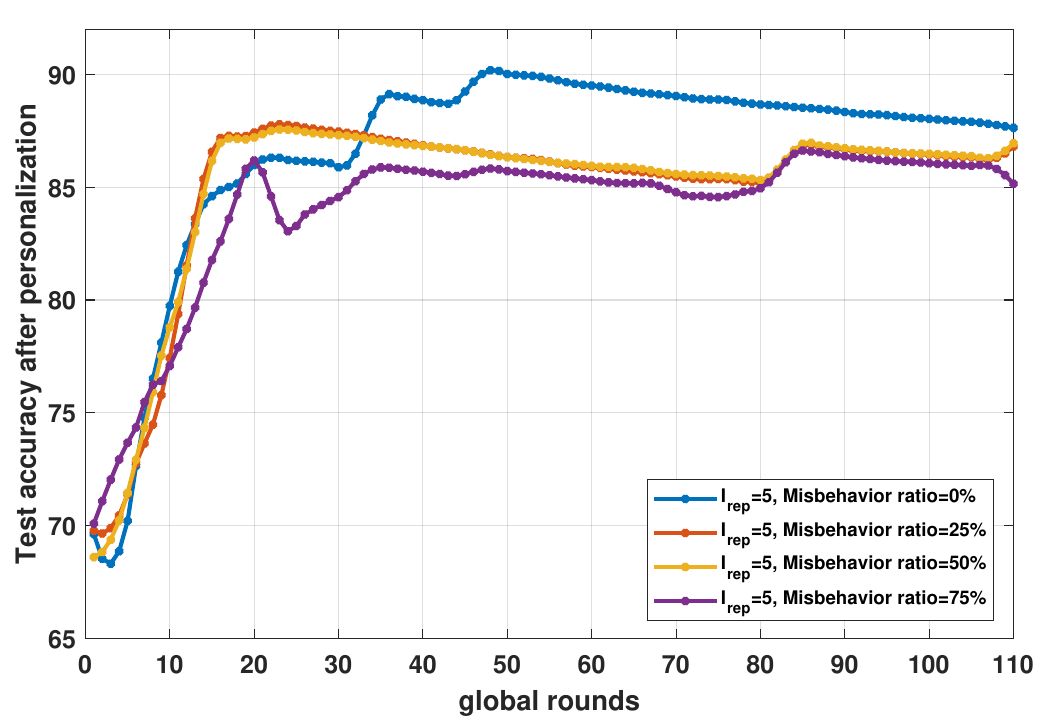}}
	 \subfigure[\label{misbehavior5_utilities}]{\includegraphics[scale=0.4]{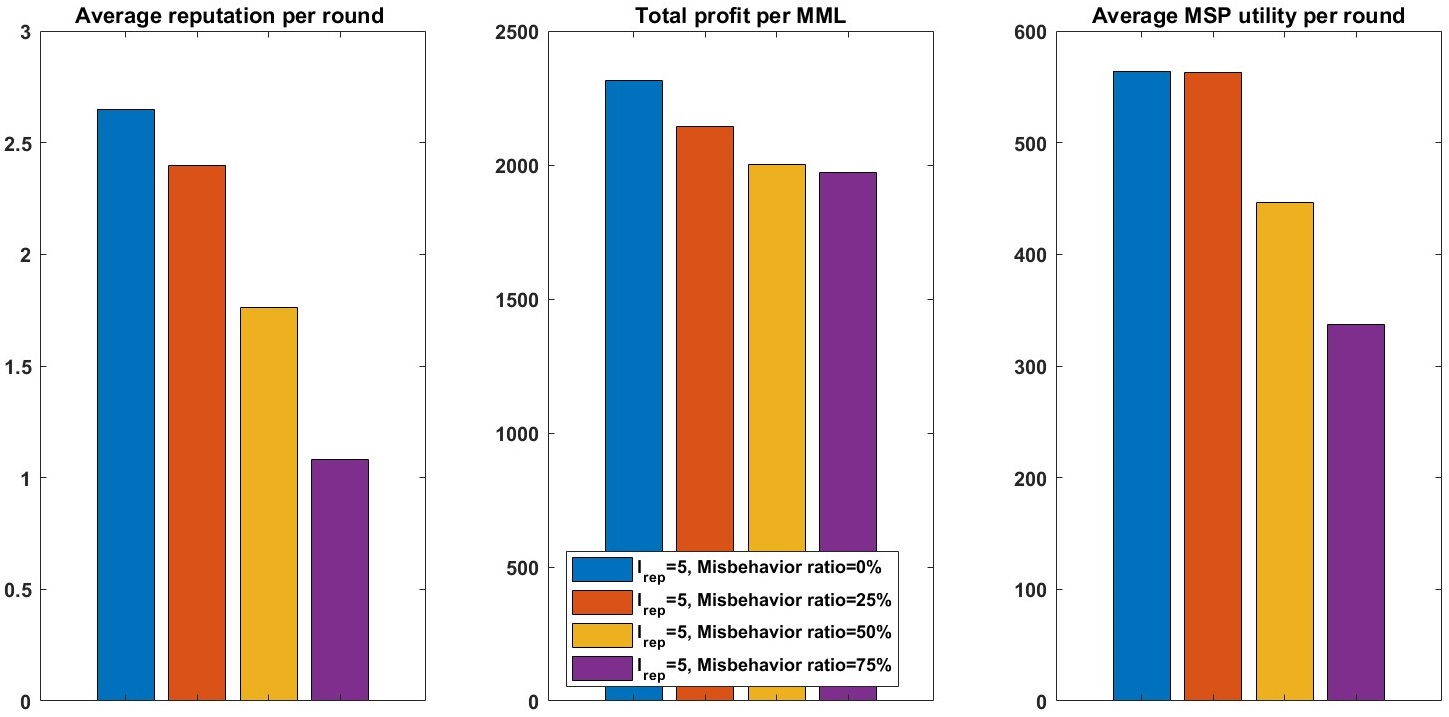}}}\\
		\mbox{
	 \subfigure[\label{misbehavior10}]{\includegraphics[scale=0.4]{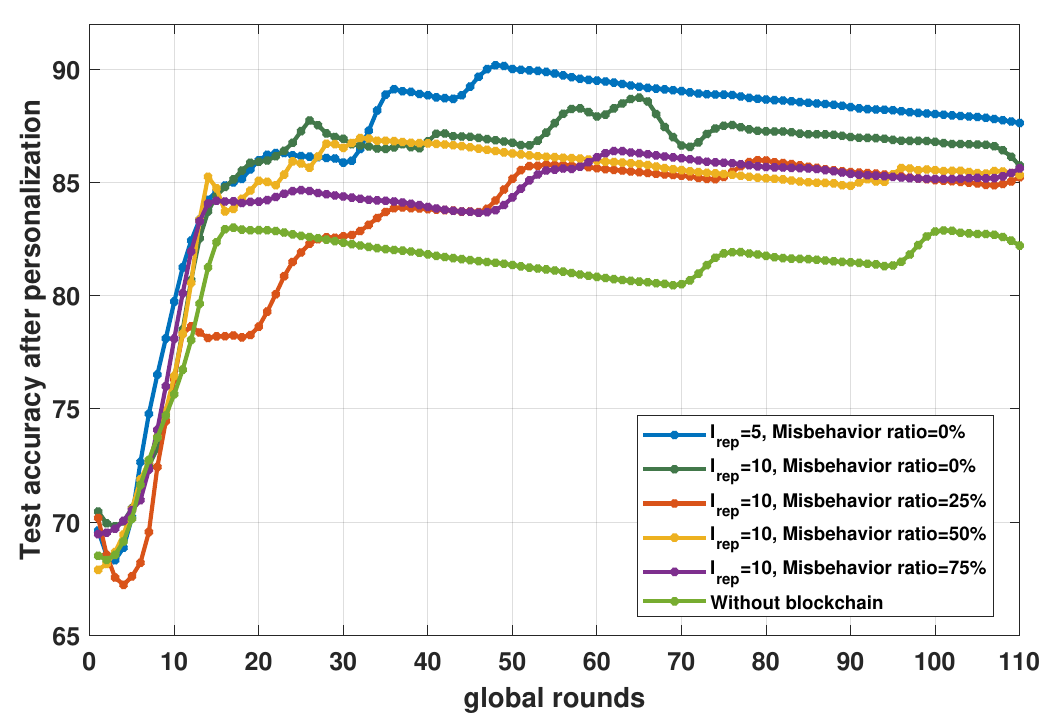}}
	 \subfigure[\label{misbehavior10_utilities}]{\includegraphics[scale=0.4]{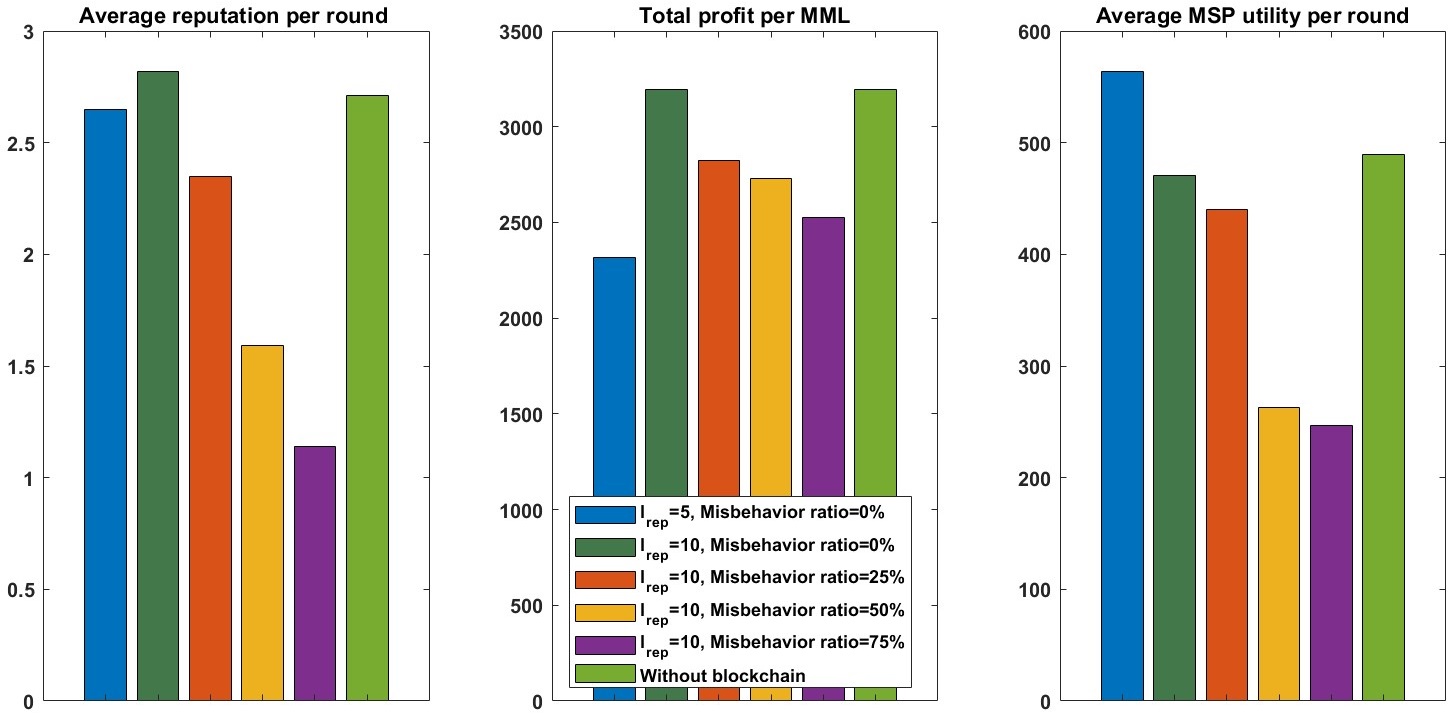}}}	
  \caption{Framework performance function of misbehavior ratio.}
   \label{misbehavior}
\end{figure*}
\subsection{Numerical analysis of the GFML performance under a non-reliable environment}
\begin{figure}[h]
\centering
	\includegraphics[scale=0.4]{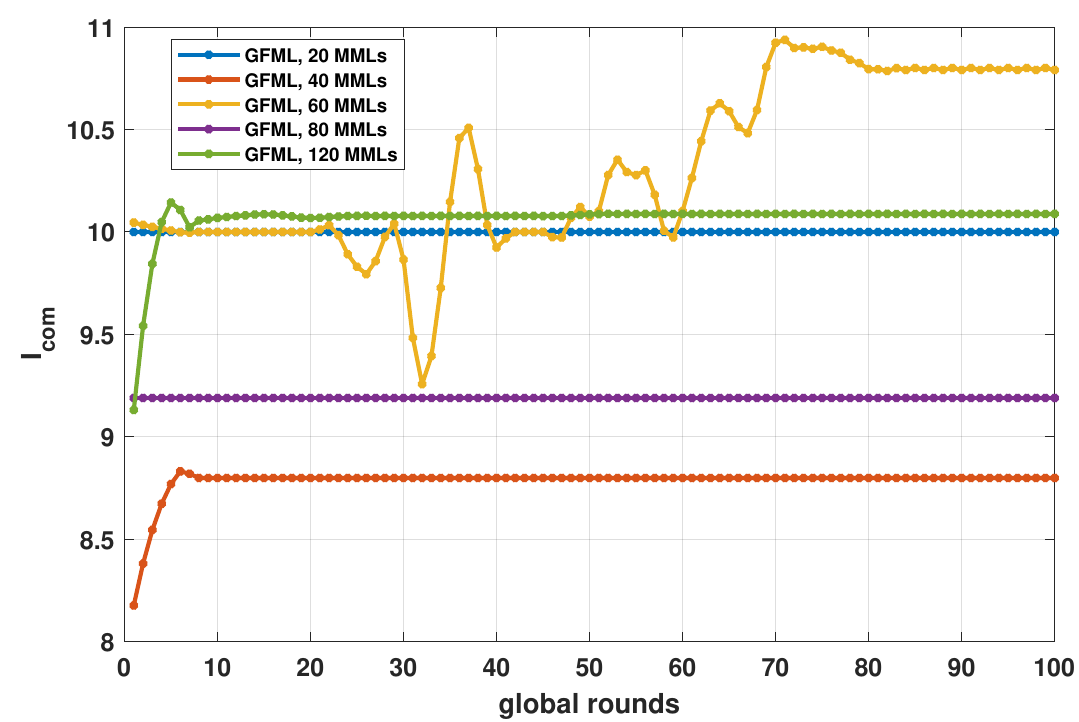}
	\caption{Optimization of $I_{com}$ throughout training rounds using Stackelberg game.}
	\label{icom}
 \vspace{-8pt}
\end{figure}

In Fig. \ref{misbehavior}, the reputation mechanism is analyzed. For that, we simulated a scenario where unreliable users do not respect the required number of local iterations $\tau$ and compute fewer epochs. We can see in Fig. \ref{misbehavior5} and Fig. \ref{misbehavior10} that the accuracy is slightly reduced when 25\%, 50\% and 75\% of MMLs misbehave, whether when $I_{rep}$ is equal to 5 or 10.  This owes to the cooperative coalition formation that balances the average reputations among coalitions, in order to have fair training circumstances under different unreliable environments. However, we notice in Fig. \ref{misbehavior10}, that the performance of the training when $I_{rep}$ is equal to 5 is slightly better than its performance when  $I_{rep}$ is equal to 10. This can be explained by the fact that the weight of the competition part of the users' utility $I^j_{com}$ is much higher when $I_{rep}$ is equal to 5 (almost double as seen in Fig. \ref{icom}). Hence, the impact of clustering by embeddings similarity is higher, which improves the training efficiency, if the misbehavior rate is 0\%. In Fig. \ref{icom}, we observe the optimization of $I^j_{com}$ throughout the training rounds using the Stackelberg game, with variation of the number of users within an honest environment. It is worth-noting that selecting $I^j_{com}$ to be much larger than $I_{rep}$ consistently results in improved system performance. However, it is essential to strike a balance, as significantly reducing the value of $I_{rep}$ can adversely affect system performance, as observed in the case of fixed coalition formation with $I_{rep}=0$. These findings highlight the need for a balance between $I^j_{com}$ and $I_{rep}$, considering resource contribution, historical reputations and potential misbehaving. Future works could explore this further.

Fig. \ref{misbehavior5_utilities} and Fig. \ref{misbehavior10_utilities} show that the payoff of the MMLs and the utility of the MSP decrease when the misbehavior rate increases, which is due to the reputation drop and the decrease of the training performance impacting both incentives and MSP profits. For the reputation scheme without blockchain, reputation values are stored in a centralized platform, vulnerable to manipulation. As illustrated in Fig. \ref{misbehavior10_utilities}, when managed by blockchain, users' reputations diminish upon falsifying their computational contributions.  In the absence of blockchain, the system experiences a decline in accuracy, yet the incentives stay high due to manipulation. Specifically,  blockchain enables a tamper-proof mechanism for managing the reputations of MMLs. Falsifying computational efforts to reduce energy consumption potentially disrupts the learning process. Our system mitigates these risks by basing reputation calculations on actual contributions to the training and timely updates with new local models. Misrepresentations are detectable through poor model performance or delayed updates, negatively affecting the user's reputation. Then, this historical reputation influences future coalition formations, ensuring that incentives for malicious behavior are minimized and guaranteeing fair training conditions across different coalitions. The incorporation of blockchain is key to this approach. Without it, data integrity could be compromised, making the system vulnerable to manipulation. Blockchain ensures that all contributions and reputations are recorded transparently and securely, preventing tampering and maintaining system performance and fairness in incentives distribution. Contrarily, in systems without blockchain, the training contributions of unreliable users are manipulated to the highest values. Blockchain ensures that such unreliable MMLs receive less incentives while preserving performance efficiency. %

Blockchain  enhances security at the expense of  scalability challenges and higher costs due to its decentralized consensus mechanisms. Despite these issues, ongoing advancements aim to balance security and scalability, and costs. Furthermore, future considerations may include using blockchain to identify and penalize low-reputation users (i.e., temporary ban from the application), promoting honesty within the framework.

\section{Conclusion}\label{section:conclusion}
Our novel framework leverages federated meta-learning to support metaverse services. Designed as a cooperative coalition formation game, it efficiently tackles statistical heterogeneity, training inefficiency, and real-time rendering. It also ensures reliable, transparent, and decentralized management of user interactions, using Blockchain technology. Blockchain users management is based on our proposed reputation mechanism, leveraging historical and potential contributions to refine and personalize user experiences. Initial evaluations demonstrate improvements in training performance by up to 10\%, task completion times by as much as 30\% over traditional clustering methods, and 5\% in training efficiency over non-blockchain systems, confirming the framework's promise in delivering tailored and trustworthy immersive experiences. Future works will consider shared embeddings privacy and incentives optimization.

\bibliography{References}
\end{document}